\documentclass[reprint,prb,graphicx,superscriptaddress]{revtex4-1}
\usepackage{bm}
\pdfoutput=1
\usepackage{graphicx}
\usepackage{amsmath}
\usepackage{subfigure}
\usepackage{array}
\usepackage{color,colortbl}
\usepackage{tabularx}
\definecolor{Gray}{gray}{0.9}
\usepackage{multirow}
\usepackage{hhline}

\begin{document}

\title{Control of surface plasmon-polaritons in magnetoelectric heterostructures}

\author{Daria O. Ignatyeva}
\affiliation{Lomonosov Moscow State University, Faculty of Physics, Leninskie Gory, Moscow, 119991, Russia}
\affiliation{Russian Quantum Center, 45 Skolkovskoye shosse, Moscow, 121353, Russia}

\author{Andrey N. Kalish}
\affiliation{Lomonosov Moscow State University, Faculty of Physics, Leninskie Gory, Moscow, 119991, Russia}
\affiliation{Russian Quantum Center, 45 Skolkovskoye shosse, Moscow, 121353, Russia}

\author{Vladimir I. Belotelov}
\affiliation{Lomonosov Moscow State University, Faculty of Physics, Leninskie Gory, Moscow, 119991, Russia}
\affiliation{Russian Quantum Center, 45 Skolkovskoye shosse, Moscow, 121353, Russia}

\author{Anatoly K. Zvezdin}
\affiliation{Russian Quantum Center, 45 Skolkovskoye shosse, Moscow, 121353, Russia}
\affiliation{A.M. Prokhorov General Physics Institute RAS, 38 Vavilov st., Moscow, 119991, Russia}
\affiliation{Moscow Institute of Physics and Technology, 9 Institutskiy per., Dolgoprudny, Moscow Region, 141700, Russia}
\affiliation{National Research University Higher School of Economics, Faculty of Physics, 21/4 Staraya Basmannaya Ulitsa, Moscow, 105066, Russia}

\author{Yujun Song}
\affiliation{Beijing Key Laboratory for Magneto-Photoelectrical Composite and Interface Science, Department of Applied Physics, Center for Modern Physics Technology, University of Science and Technology Beijing, Beijing, 100083, China}

\begin{abstract}
The work is devoted to the investigation of the surface plasmon polaritons in the metal - dielectric heterostructures containing materials with magnetoelectric properties. The analysis of various possible configurations shows that the magnetoelectric effect influences the plasmonic surface wave polarization, localization and dispersion. The essential case of chromium (III) oxide as a magnetoelectric dielectric is described in detail, revealing the switching between different regimes of magnetoelectric impact on plasmon-polaritons. Plasmonic heterostructures with magnetoelectric constituents can be used as a polarization- and dispersion-sensitive instrument to reveal the magnetoelectric coupling and besides might act as an optical tool for control of the surface plasmon-polariton dispersion and dynamics via external electric and magnetic fields. Apart from that, the described effects can be applied to the manipulation of the far-field optical response of plasmonic structures. Composite magnetoelectric materials like $\mathrm{BaTiO_3/BiFeO_3}$ that exhibit giant magnetoelectric effect are very promising for these purposes.
\end{abstract}


\maketitle

\section{Introduction}

Nowadays, control of surface plasmon polaritons (SPP) by an external stimulus is of prime importance for telecommunication and other applications and is usually referred to as active plasmonics~\cite{MacDonald:2009,Strohfeldt:2014,Papaioannou:2012,Si:2014,Cetin:2012,shalaev}. The active plasmonics utilizes strong dependence of the SPP properties, namely its dispersion, localization, the transverse profile and polarization on the permittivity of the surrounding media. Slight variations of the dielectric permittivity result in a significant shift of the SPP resonance which is widely used nowadays for sensing applications~\cite{Homola:1999, piliarik2009surface, willets2007localized, Ignatyeva:2016} and all-optical light control~\cite{kar2017long, chen2011highly, lu2011ultrafast, rotenberg2010ultrafast, kreilkamp2016terahertz, ignatyeva:2012:APA, ignatyeva2014femtosecond}. The impact of magnetooptical gyrotropy on the SPP properties is promising for ultrafast modulation of the laser radiation~\cite{Temnov:2010,shcherbakov2014femtosecond,Kalish:2014,Belotelov:2014,belotelov2009giant}, while the presense of optical activity~\cite{Ignatyeva:2012:PRA,Ignatyeva:2011:THZ} performs the modification of the SPP polarization. It was shown that such issues of magnetoelectric coupling as chirality~\cite{Guangcan:2014} and the axion effect in topological insulators~\cite{Ignatyeva:2017} can be revealed in plasmonic structures via observation of the changes in the SPP excitation conditions or polarization mediated by them.

The presence of the magnetoelectric (ME) properties of the SPP sustaining structures is very promising in the context of the SPP control. The ME effect might modify SPP properties, thus providing another tool for light manipulation. In this paper we address this problem. Although surface electromagnetic waves at the interfaces of bianisotropic dielectrics have been studied earlier~\cite{tarkhanyan2011nonradiative, galynsky2004integral}, the properties of the SPP waves supported by metal-magnetoelectric structures were not addressed before. At the same time, compared to the surface electromagnetic waves in dielectrics, SPP waves are characterized by stronger electromagnetic field concentration in a region of about 100 nm and higher sensitivity to the medium properties which makes them promising for efficient magnetoelectric light control.

In Sec.II of this paper we provide a theory of SPP waves in a most general case of a metal/magnetoelectric interface. We provide the description and classification of different magnetoelectric configurations and phases corresponding to different orientations of magnetoelectric crystals and different types of magnetoelectrics which makes it possible to predict the variations of SPP characteristics in the particular structure only by the form of its magnetoelectric tensor. In Sec.III of the paper we apply our theory to $\mathrm{Cr_2O_3}$ crystal which not only has the magnetoelectric response but also provides the possibility to switch it due to the spin-flop effect. We analytically and numerically investigate the SPP modes in several basic configurations. We show how laser radiation can be modulated in the far-field using SPPs in magnetoelectric structures. Although the effects found in $\mathrm{Cr_2O_3}$ are rather moderate, similar approach can be applied to other magnetoelectrics exhibiting stronger ME response. For example, artificial metamaterials and composite multilayered structures like $\mathrm{BaTiO_3/CoFe_2O_4}$ or $\mathrm{BaTiO_3/BiFeO_3}$ are designed to provide the enhancement of the ME effects~\cite{Shuvaev:2011,Zheludev:2012, popkov2016origin, bichurin2011magnetoelectricity, sreenivasulu2012magnetoelectric}. The structure that is characterized by the typical value of the magnetoelectric constant of 0.1 has recently been designed~\cite{Lorenz:2015}, that makes the optical ME effects of practical interest.

The magnetoelectric (ME) effect is a coupling between the electric and magnetic field components so that the magnetic field induces electric polarization and the electric field induces magnetization in a material. It can be observed in antiferromagnetics, multiferroics, and topological insulators~\cite{Eerenstein:2006, Pisarev:1994, Fiebig:2005,Achida:2014} as well as in some artificial metamaterials~\cite{Shuvaev:2011}. The following constitutive equations describe the ME effect at optical frequencies (permeability is set to unity)~\cite{bichurin2011magnetoelectricity}:
\begin{eqnarray}
\mathbf{D} = \hat{\varepsilon}^0\mathbf{E} + \hat{\alpha} \mathbf{B} \label{eqn_econstitutive}\\
\mathbf{H} = \mathbf{B} - \hat{\alpha}^+ \mathbf{E}
\label{eqn_hconstitutive}
\end{eqnarray}
where $\hat{\alpha}$ is the ME susceptibility tensor whose symmetry and characteristic component values depend on the material. Note that the energy conservation law requires the ME susceptibility tensor $\hat{\alpha}$ to be the same in (\ref{eqn_econstitutive}) and (\ref{eqn_hconstitutive}) for all of the non-absorbing materials. We restrict our consideration to the case of real components of $\hat \alpha$.

The case of the isotropic form of the ME susceptibility tensor is similar to the so-called axion effect~\cite{Wilczek:1987,Li:2010,Hasan:2010}. The latter effect may appear if a perturbation breaking time reversal symmetry is present. The axion effect can be described by the effective Lagrangian density $L_{\theta} = \left[ \theta e^2 / (4 \pi^2 \hbar c ) \right](\mathbf{E}\cdot \mathbf{B})$ where $\theta$ is the quantized axion angle that is an odd integer multiple of $\pi$. The axion itself is a hypothetical elementary particle that is a candidate for dark matter~\cite{Preskill:1983}.

The non-diagonal ME tensor corresponds to the other type of ME effect that is the toroidal one~\cite{Popov:1999}. The latter is related to the non-zero toroidal moment. As it has been shown recently, one of the reasons for the fundamental interest in this area lies in the fact that toroidal structures allow realization of non-radiating sources which give way to the time-dependent Aharonov-Bohm effect and corresponding novel effects and applications~\cite{Zheludev:2013, Zheludev:2016}.

One of the prominent representatives of the ME materials is chromium (III) oxide  $\mathrm{Cr_2O_3}$~\cite{Obukhov:2008}. Its symmetry leads to the following form of the ME susceptibility tensor~\cite{Popov:1999}:
\begin{equation}
\hat \alpha = \begin{pmatrix}
\lambda_1 L_{y'} + \lambda_3 L_{z'} & \lambda_1 L_{x'} &\lambda_2 L_{x'} \\
\lambda_1 L_{x'} & - \lambda_1 L_{y'} + \lambda_3 L_{z'} & \lambda_2 L_{y'} \\
\lambda_4 L_{x'} & \lambda_4 L_{y'} & \lambda_5 L_{z'} \end{pmatrix},
\label{alpha_in_cr2o3}
\end{equation}
where $\mathbf{L}$ is the unitary antiferromagnetic vector and $\lambda_j$ are the coefficients that for the selected material have the typical values of the order of $10^{-4}$ ~\cite{Wiegelmann:1994}. This magnetoelectric susceptibility tensor is written in the local coordinate system associated with the crystal symmetry: $x'$-axis is parallel to the two fold axis and $z'$-axis is parallel to the three fold symmetry axis (the $c$-axis)~\cite{Krichevtsov:1993}.

An important feature of $\mathrm{Cr_2O_3}$ is that its magnetoelectric response can be switched between two types. Such switching is possible due to the "spin-flop" effect that is a reorientation of the spins in strong magnetic fields~\cite{Foner:1963} exceeding the threshold value of about 7 T. Therefore, below the Neel temperature of $T_N=307\text{~K}$ the symmetry of the ME tensor is quite different in the two phases, below and above the spin-flop. Below the threshold, in so-called low-field phase, the spins and the antiferromagnetic vector of $\mathrm{Cr_2O_3}$ both are oriented along the $c$-axis. At this the ME susceptibility tensor has the isotropic form, so that the electrodynamics demonstrates the axion-like behaviour.

In the strong magnetic fields applied along the three-fold axis exceeding this threshold value the phase transition to the spin-flop phase occurs. It results in the $90^\circ$ rotation of the antiferromagnetic vector that becomes perpendicular to the $c$-axis~\cite{Wiegelmann:1994,Fiebig:1996}. The ME tensor in this case has non-diagonal components that affect the electrodynamics of the crystal in a different way compared to the low-field phase.

Therefore, the spin-flop transition is a transition from axion-like to the toroidal magnetoelectricity. Due to that $\mathrm{Cr_2O_3}$ can be considered as a unique object on which one can trace the similarities and difference between the axion-like and toroidal effects that is an important fundamental problem. So $\mathrm{Cr_2O_3}$ represents an appropriate model object for the study of optical ME responce. 

Due to the switchable ME effect the chromium (III) oxide $\mathrm{Cr_2O_3}$ is perspective for active plasmonics. Although the changes in the ME susceptibility tensor produced by spin-flop transition are rather small in comparison with anisotropy of the $\mathrm{Cr_2O_3}$ crystal that is $\Delta n = 5.8 \cdot 10^{-2}$ it might be possible to reveal both the ME effect and the spin-flop phase transition by the surface plasmon polariton resonance which is sensitive to the ME properties. Therefore, the plasmonic surface of a $\mathrm{Cr_2O_3}$ crystal coated with a metal, e.g. gold, provides an optical tool for investigation of the $\mathrm{Cr_2O_3}$ phase transitions. Moreover, a perspective feature of the ME plasmonic interfaces is that one could get use of their ME tunability at static fields. Namely, optical resonances might be modified by application of a constant external electric field. $\mathrm{GaFeO_3}$ material is one of the most promising for that due to the presence of the toroidal moment~\cite{popov1997linear}. It makes such materials of interest for applications.

\section{Theoretical approach to the magnetoelectric surface plasmon polaritons}
\subsection{General description of the ME SPPs}
For the convenience we associate the coordinate system with the plasmonic interface so that the SPP propagates along $Ox$ axis and $Oz$ axis is perpendicular to the interface (Fig.~\ref{fig_geometrya}). For the most general description of the considered system the electromagnetic field of the SPP in a ME dielectric is represented as the sum of the two partial TE and TM components which can have different polarization ($\mathbf{E}_j$) and localization ($\gamma_j$), in general, but equal propagation constants $\beta$:
\begin{equation}
\mathbf{E}_{spp} (z>0) = \left( \mathbf{E}_{TM} e^{-\gamma_{TM} k_0 z} + \mathbf{E}_{TE} e^{-\gamma_{TE} k_0  z} \right) e^{i \beta k_0  x},
\label{e_spp_cr}
\end{equation}
while in the metal it has the conventional form:
\begin{equation}
\mathbf{E}_{spp} (z<0)= \mathbf{E}_{m} e^{\gamma_{m} k_0  z} e^{i \beta k_0   x},
\label{e_spp_m}
\end{equation}
where $\mathbf{E}_{m}$ determines the polarization vector and $\gamma_{m}$ is responsible for the SPP localization in the metal. The localization coefficients as well as the propagation constant are normalized by vacuum wavenumber $k_0$. Once the electromagnetic field is represented in accordance to Eqs.~(\ref{e_spp_cr}) and (\ref{e_spp_m}), we can solve Maxwell's equations with the corresponding boundary conditions and constitutive equations~(\ref{eqn_econstitutive}), (\ref{eqn_hconstitutive}). Accurate analysis shows that in the absence of the surface charge and the surface current the boundary conditions take the conventional forms. This means that the normal components of $\mathbf{D}$ and $\mathbf{B}$, as well as the tangential components of $\mathbf{E}$ and $\mathbf{H}$ are continuous over the interface.

The ME susceptibility tensor $\hat \alpha$ can be decomposed into the symmetric and antisymmetric parts: $\hat \alpha = \hat \alpha^S + \hat \alpha^{AS}$. The symmetric part of the ME susceptibility tensor is responsible for the so-called axion coupling while the antisymmetric one is associated with the toroidal moment that characterizes the specific kind of magnetic ordering~\cite{Schmid:1973,Popov:1999,gusev2014surface}. The antisymmetric part can be expressed in terms of characteristic vector $\boldsymbol{\tau}$ as
\begin{equation}
\alpha_{ij}^{AS} = \epsilon_{ikj} \tau_k,
\label{alpha_tensor}
\end{equation}
where $\epsilon_{ijk}$ is Levi-Civita symbol, and vector $\boldsymbol{\tau}$ is proportional to the toroidal moment vector $\mathbf{T}$ that is dual to $\hat \alpha^{AS}$.

Using the plane-wave representation for the SPP in the ME dielectric with complex wave vector $\boldsymbol{\kappa}^{(j)} = \{ \beta; 0; i\gamma_j\}$ and using Maxwell's equations, the consitutive equations (\ref{eqn_econstitutive}) and (\ref{eqn_hconstitutive}) and Eq.~(\ref{alpha_tensor}) one can obtain the following wave equation:
\begin{equation}
\left[\boldsymbol{\kappa} \times \left[\boldsymbol{\kappa} \times \mathbf{E}\right]\right] + k_0^2 \hat \varepsilon(\boldsymbol{\kappa}) \mathbf{E} = 0,
\end{equation}
where $k_0$ is the vacuum wavenumber and the effective dielectric permittivity tensor $\hat \varepsilon(\boldsymbol{\kappa})$ that includes ME contribution has the form:
\begin{equation}
\varepsilon_{ij}(\boldsymbol{\kappa}) = \varepsilon_{ij}^0 -  \left(\epsilon_{isj} \kappa_s (\alpha_j - \alpha_i) + \kappa_j \tau_i + \kappa_i \tau_j - 2 (\boldsymbol\kappa \boldsymbol\tau) \delta_{ij}\right),
\end{equation}
where $\alpha_i$ are the diagonal components of the magnetoelectric tensor and $\delta_{ij}$ is Kronecker delta. Here the linear in $\hat \alpha$ approximation is applied.

This equation is rather complicated, especially in the case of $\mathrm{Cr_2O_3}$ that has anisotropic tensor $\hat \varepsilon^{0}$ and both symmetric and antisymmetric components of the ME susceptibility tensor. In order to reveal the impact of the symmetric and antisymmetric parts of the ME susceptibility tensor on the SPP we first address the idealized problems of isotropic media with fully symmetric or fully antisymmetric ME
tensors.

\subsection{Axion-type impact on SPPs}

Here we consider an isotropic dielectric possessing the dielectric permittivity $\varepsilon^0$ with a diagonal ME susceptibility tensor in the form $\hat \alpha = \alpha \hat I$, where $\hat I$ is the identity tensor.

Although the constitutive equations differ from the case of usual dielectric, the bulk electromagnetic wave does not experience the influence of the axion effect. The following wave equation describing bulk wave propagation can be obtained directly from Maxwell's equations for the medium with the axion effect:
\begin{equation}
\left[\nabla \times \left[\nabla \times \mathbf{E}\right]\right] + \frac{\varepsilon^0}{c^2} \frac{\partial^2 \mathbf{E}} {\partial t^2} = 0.
\label{wave_eqv_axion}
\end{equation}
Since this equation does not contain any axion terms the bulk wave keeps its polarization and the propagation constant is the same for any propagation directions and any polarization states. The only change that comes from the axion effect is the orthogonality of the vectors: $\mathbf{E} \perp \mathbf{B}$ and $\mathbf{D} \perp \mathbf{H}$ while vectors $\mathbf{E}$ and $\mathbf{D}$, as well as $\mathbf{B}$ and $\mathbf{H}$ are non-collinear.

The impact of the axion effect on the electromagnetic wave properties reveals itself at the interface between axion and non-axion medium. This can be illustrated via the conventional boundary conditions between a medium-1 and a medium-2 that in the presence of the axion effect in the medium-2 have the form:
\begin{eqnarray}
D_{n}^{(1)} - D_{n}^{(2)} = \varepsilon^{(1)} E_n^{(1)}  - \varepsilon^{(2)} E_n^{(2)} - \alpha B_n^{(2)} = 0, \\
H_{\tau}^{(1)} - H_{\tau}^{(2)} = B_\tau^{(1)} - B_\tau^{(2)} + \alpha E_\tau^{(2)} =  0.
\end{eqnarray}
The latter terms can be treated as "effective" axion surface charges with density $ \rho_{ax} = (1/4 \pi)\alpha B_n^{(2)}$ and surface currents $j_{ax} = (c/4 \pi) \alpha E_\tau^{(2)}$.

The peculiarities of the transmission or reflection of bulk electromagnetic waves in structures with axion/non-axion layers were considered earlier~\cite{Wang-Kong:2010, Chang:2009, Liu:2014}. In this paper we investigate surface plasmons at the axion-dielectric/metal interface and their influence on the optical reflection.

Using the linear-in-$\alpha$ approximation and neglecting all the terms that are associated with powers of $\alpha$ higher than first, we can fully describe the SPP. The dispersion has no linear-in-$\alpha$ terms so the axion effect does not noticeably change the propagation constant and the surface plasmon resonance frequency.

Equation~(\ref{wave_eqv_axion}) implies that there is a polarization degeneracy of the bulk modes in a medium with the axion effect. Therefore partial TM and TE components of the SPP wave (see Eq.(\ref{e_spp_cr})) should have equal localization coefficients $\gamma_{TM}= \gamma_{TE} = \gamma_{ax}$. Notice that due to the axion coupling between the TM and TE components they are both present in the SPP field. The analytical solution of the SPP eigenmode problem taking $H_y=1$ gives the following expressions for the electromagnetic fields in axion media:
\begin{eqnarray}
\mathbf{E}_{ax} = \begin{pmatrix} i \frac{\gamma_{ax}}{\varepsilon^{0}} \\ -\alpha \frac{1}{\varepsilon^0 - \varepsilon_m}\\ - \frac{\beta}{\varepsilon^{0}} \end{pmatrix} , ~~
\mathbf{H}_{ax} = \begin{pmatrix}  i\alpha \frac{\gamma_{ax}\varepsilon_m}{\varepsilon^{0}(\varepsilon^0 - \varepsilon_m)} \\ 1 \\ - \alpha \beta \frac{\varepsilon_m}{\varepsilon^{0}(\varepsilon^0 - \varepsilon_m)}\end{pmatrix},
\label{e_h_axion}
\end{eqnarray}
where $\beta$ is the SPP propagation constant. Let us define the ratio between TE and TM components as $E_y/H_y$ ratio that in the case of the axion effect is:
\begin{equation}
\left| \frac{E_y}{H_y} \right| = \frac{ \alpha}{\varepsilon^{0} - \varepsilon_m}
\label{eh_axion}
\end{equation}
The phases of $E_y$ and $E_z$ components coincide so that the polarization ellipse of the SPP wave that in the absence of the axion effect was oriented in $Oxz$ plane experiences the slight rotation around $Ox$ axis by an angle of $E_y / E_z$ due to the axion effect. While the ME susceptibility tensor is isotropic, the direction of this rotation is strictly defined by the SPP propagation direction, since the angle is odd in $\beta$. This may be referred as the locking between the SPP propagation direction and the orientation of the polarization ellipse, similarly to the spin-momentum locking for electronic surface states.

We note that the provided analysis of the SPPs at the interface between an isotropic axion medium and a metal can be directly applied to the topological insulators~\cite{Ignatyeva:2017} with $\alpha=1/137$ that possess the axion properties due to the perturbations of the time-reversal symmetry~\cite{Hasan:2010,Li:2010}.

\subsection{Impact of toroidal magnetic moment on SPPs}
Let us consider the interface between a dielectric possessing toroidal moment $\mathbf{T}$ and a metal. Taking into account Eqs.~(\ref{eqn_econstitutive}), (\ref{eqn_hconstitutive}) and (\ref{alpha_tensor}) and applying linear in $\boldsymbol\tau$ approximation the constitutive equations for the dielectric can be rewritten in the following way:
\begin{eqnarray}
\mathbf{D} = {\varepsilon}^0\mathbf{E} + [\boldsymbol\tau \times \mathbf{H}]\\
\mathbf{B} = \mathbf{H} - [\boldsymbol\tau \times \mathbf{E}]
\end{eqnarray}

The presence of the toroidal moment leads to the effect of non-reciprocal wave propagation, so that the waves propagating along $\boldsymbol\tau$ and opposite to $\boldsymbol\tau$ possess different wavenumbers. The relations between $\mathbf{E}$ and $\mathbf{H}$ vectors are also modified and, moreover, $\mathbf{E}$ and $\mathbf{H}$ vectors are generally no longer perpendicular to the wavevector~\cite{Kalish:2007}. The essential point is that the impact of the toroidal moment on the bulk wave properties is polarization-independent, similarly to the case of the axion effect.

The properties of SPPs in media with toroidal moment are strongly dependent on the direction of $\boldsymbol\tau$. There are three basic configurations:
\begin{enumerate}
\item the longitudinal configuration: $\boldsymbol\tau$ lies in the interface plane along the SPP propagation direction ($\boldsymbol\tau$ is directed along $Ox$ axis);
\item the transversal configuration: $\boldsymbol\tau$ lies in the interface plane perpendicular to the SPP propagation direction ($\boldsymbol\tau$ is directed along $Oy$ axis);
\item the polar configuration: $\boldsymbol\tau$ is perpendicular to the interface plane ($\boldsymbol\tau$ is directed along $Oz$ axis).
\end{enumerate}

In the longitudinal configuration both propagation constant and localization coefficients are affected by the toroidal moment:
\begin{eqnarray}
\beta = \sqrt{\frac{\varepsilon_m \varepsilon^0}{\varepsilon_m + \varepsilon^0}} - \tau \frac{\varepsilon_m^2 }{\varepsilon_m^2 - (\varepsilon^0)^2} \label{nr-case1}\\
\gamma_{TM} =  \frac{\varepsilon^0}{\sqrt{-(\varepsilon_m + \varepsilon^0)}} - \tau \frac{\varepsilon^0 \sqrt{-\varepsilon_m \varepsilon^0 }}{\varepsilon_m^2 - (\varepsilon^0)^2} \\
\gamma_m =  \frac{-\varepsilon_m}{\sqrt{-(\varepsilon_m + \varepsilon^0)}} + \tau \frac{\varepsilon_m \sqrt{-\varepsilon_m \varepsilon^0 }}{\varepsilon_m^2 - (\varepsilon^0)^2}
\end{eqnarray}

Equation~(\ref{nr-case1}) demonstrates the nonreciprocal propagation of the SPP. Indeed, the SPPs propagating along $\boldsymbol\tau$ and opposite to $\boldsymbol\tau$ possess different propagation constants.
The polarization of the SPP remains pure TM, so that $\mathbf{E}_{TE}=0$. However, the relations between the field components inside the dielectric are slightly modified. Taking $H_y=1$ they are the following:
\begin{equation}
\mathbf{E} = - \begin{pmatrix} i \frac{1}{\sqrt{-(\varepsilon_m + \varepsilon^0)}} - i \tau \frac{\sqrt{-\varepsilon_m \varepsilon^0 }}{\varepsilon_m^2 - (\varepsilon^0)^2} \\ 0 \\ \sqrt{\frac{\varepsilon_m }{\varepsilon^0 (\varepsilon_m + \varepsilon^0)}} - \tau \frac{\varepsilon^0 }{\varepsilon_m^2 - (\varepsilon^0)^2}  \end{pmatrix}, ~~\mathbf{H} = \begin{pmatrix} 0 \\ 1 \\ 0 \end{pmatrix} .
\label{tor_case1}
\end{equation}

It follows from Eq.~(\ref{tor_case1}) that if the SPP propagates along $\mathbf{T}$ the electric field components diminish and, moreover, the polarization ellipse becomes narrower as its semiaxes ratio $E_x/E_z$ gets smaller. At the same time, the semiaxes ratio of the polarization ellipse for the metal $E_z/E_x$ also diminishes. For the opposite direction of the SPP propagation the results are opposite.

In the polar configuration only localization constant for the dielectric changes upon the toroidal moment:
\begin{equation}
\gamma_{TM} =  \frac{\varepsilon^0}{\sqrt{-(\varepsilon_m + \varepsilon^0)}} - i \tau
\end{equation}

Unlike the case of the longitudinal configuration, the term related to the magneto-electric effect is imaginary. It means that the wavevector has a real normal component. However, this component is small in comparison with the imaginary normal component and the tangential one, so that the energy leaking is also small.

Note that in this case the SPP propagation is reciprocal. It is quite obvious since the configuration is isotropic in the interface plane.

The field components remain unchanged compared to the non-magnetoelectric case.

On the contrary, in the transversal configuration the propagation constant and the localization coefficients are independent of the toroidal moment. At the same time, the electromagnetic field acquires TE components, changing the mode polarization. Due to the optical isotropy $\gamma_{TM} = \gamma_{TE}$. The field inside the dielectric has a form:
\begin{equation}
\mathbf{E} = - \frac{1}{\varepsilon^0}\begin{pmatrix} i \gamma_{TM} \\ i \tau \frac{\beta}{\gamma_{TM} + \gamma_m} \\ \beta  \end{pmatrix}, ~~\mathbf{H} = \begin{pmatrix} -\tau \frac{\gamma_m \beta}{\varepsilon^0 (\gamma_{TM} + \gamma_m)} \\ 1 \\ i \tau \frac{\gamma_{TM}\gamma_m - \varepsilon^0}{\varepsilon^0 (\gamma_{TM} + \gamma_m)} \end{pmatrix},
\label{tor_case2}
\end{equation}
so the TE/TM ratio is
\begin{equation}
\frac{E_y}{H_y} = - i \tau \sqrt{-\frac{\varepsilon_m}{\varepsilon^0}}\frac{1}{\varepsilon^0 - \varepsilon_m}.
\end{equation}

Unlike the case of the axion effect, $E_x$ and $E_y$ components are in-phase, so the plane of the polarization ellipse turns around $Oz$ axis by the angle of $E_y/E_x$ compared to the non-toroidal case. The magnetic field acquires ellipticity with a semiaxes ratio of $|H_z/H_y|$ and experiences the turn by an angle of $H_x/H_y$.
Note that these effects are linear in $\mathbf{T}$, so that upon the toroidal moment reversal the polarization ellipse turns in the other direction and the sign of $H_z$ component also changes.

The results of Sections IIB and IIC are summarized in Table~\ref{tabular:idealized}.
\renewcommand{\arraystretch}{1.3} 
\begin{table}
\caption{The ME impact on the SPP modes in idealized cases.}
\label{tabular:idealized}
\begin{tabular}{|>{\centering\arraybackslash}p{0.18\columnwidth}|>{\centering\arraybackslash}p{0.18\columnwidth}|>{\centering\arraybackslash}p{0.18\columnwidth}|>{\centering\arraybackslash}p{0.18\columnwidth}|>{\centering\arraybackslash}p{0.18\columnwidth}|}
\hline
\rowcolor{Gray}[.85\tabcolsep]
\multicolumn{2}{|c|}{~} & Propagation constant & Localization depth & Polarization \\
\hline
\multicolumn{2}{|c|}{The axion effect}& no change & no change & $E_y/H_y$ is real, the polarization ellipse is tilted around $Ox$ axis \\
\hline
\multirow{3}{0.2\columnwidth}{\centering The toroidal effect} & $\mathbf{T}\parallel$ interface, SPP $\parallel\mathbf{T}$ & linear in $\mathbf{T}$, the SPP propagation is non-reciprocal & changes in both media & the eccentricity of the polarization ellipse changes \\ \cline{2-5}
&  $\mathbf{T}\parallel$ interface, SPP $\perp\mathbf{T}$ & no change & no change & $E_y/H_y$ is imaginary, the polarization ellipse is tilted around $Oz$ axis \\ \cline{2-5}
& $\mathbf{T}\perp$ interface& no change & changes in dielectric &no change \\   \hline
\end{tabular}
\end{table}

\section{Surface plasmon polaritons at the metal/$\mathrm{Cr_2O_3}$ interface: optical anisotropy and magnetoelectricity}
\subsection{General features of the SPP at the metal/$\mathrm{Cr_2O_3}$ interface}

\begin{figure}
\centering
\subfigure[]{\includegraphics[width = 0.38\columnwidth]{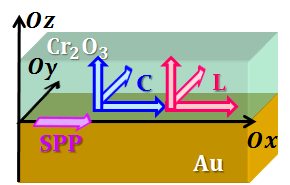} \label{fig_geometrya}}
\subfigure[]{\includegraphics[width = 0.58\columnwidth]{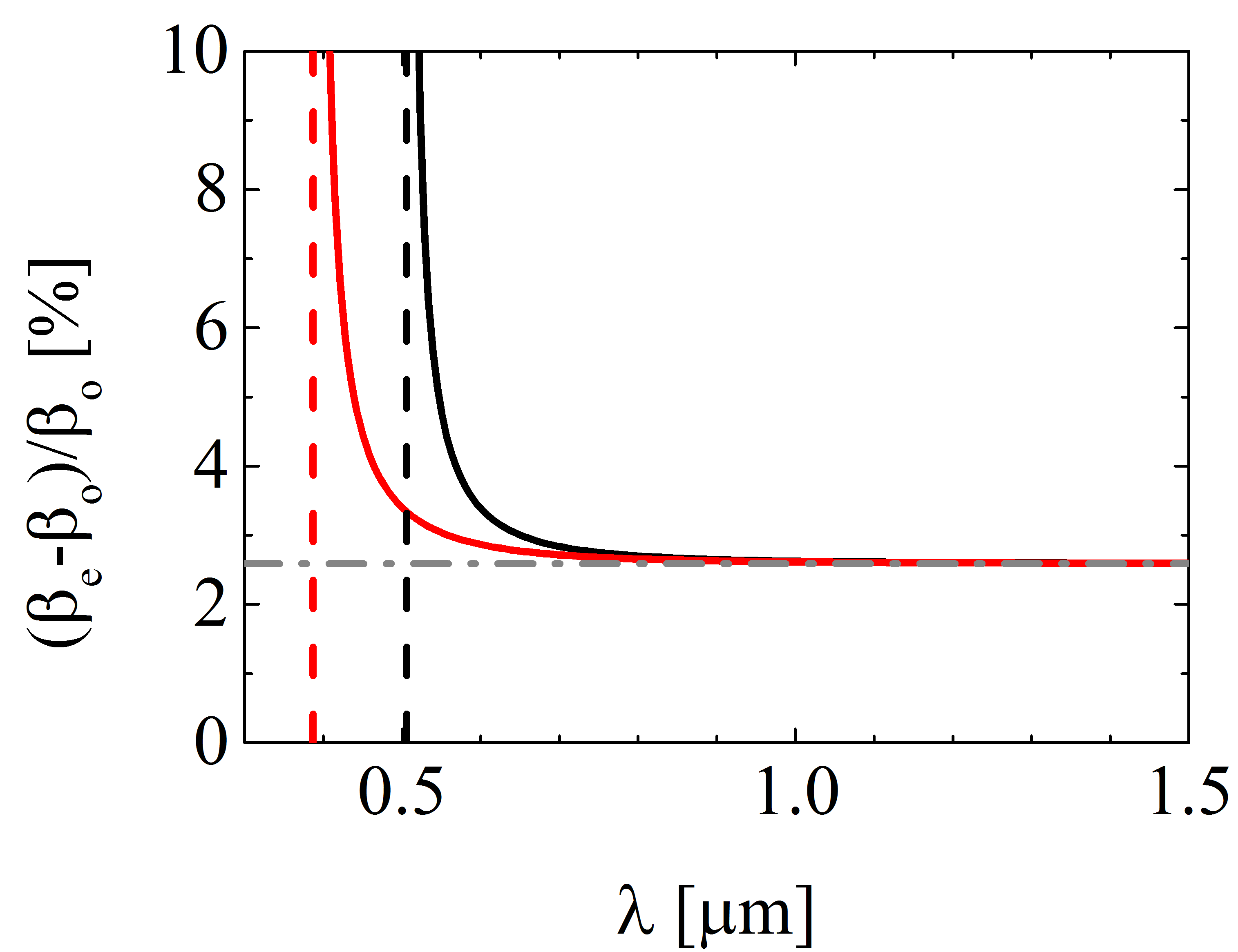}}
\caption{(a) Possible configurations of the SPP propagation at the metal/$\mathrm{Cr_2O_3}$ interface. (b) The birefringence of the SPP modes at $\mathrm{Cr_2O_3}$ -- gold (black solid line) or silver (red solid line) interfaces. Vertical dashed lines correspond to surface plasmon frequencies $\omega_{sp}$ while gray dash-dot line shows the birefrigence asymptote.}
\label{fig_geometry}
\end{figure}

The observed impact of the ME effect in a $\mathrm{Cr_2O_3}$ crystal on the SPP modes depends significantly on the relative orientation of the crystal $c$-axis and the antiferromagnetic vector $\mathbf{L}$ with respect to the plasmonic interface between a metal and the $\mathrm{Cr_2O_3}$ crystal as well as with respect to the direction of the SPP propagation. For the detailed analysis we choose 9 basic possible configurations of the SPP propagation along the interface (see Fig.~\ref{fig_geometry}a) which correspond to 3 possible orientations of the {\it c}-axis and 3 possible orientations of $\mathbf{L}$ vector with respect to the propagation direction.

First of all, we should mention that $\mathrm{Cr_2O_3}$ crystal possesses optical anisotropy, for the wavelength of 633~nm $\varepsilon_{\perp}=4.9$ and $\varepsilon_{\parallel}=5.15$ \cite{Krichevtsov:1993}. Therefore the dispersion of the SPP waves at its interface depends on its orientation. If the $c$-axis is oriented parallel to the interface and perpendicular to the SPP direction the surface wave can be called "ordinary" since it does not experience anisotropy and its propagation constant is determined as:
\begin{equation}
\beta_o = \sqrt{\frac{\varepsilon_{\perp} \varepsilon_m}{\varepsilon_{\perp} + \varepsilon_m}},
\label{beta_o}
\end{equation}
where $\varepsilon_{\perp}$ is the component of the dielectric permittivity tensor corresponding to the perpendicular to the $c$-axis direction ($\varepsilon_{\parallel}$ is the component parallel to the $c$-axis) and $\varepsilon_m$ is the permittivity of a metal.

In the other cases optical anisotropy affects the SPP and its "extraordinary" dispersion takes the following forms:
\begin{equation}
\beta_e = \sqrt{\frac{\varepsilon_m \varepsilon_{\parallel} (\varepsilon_m - \varepsilon_{\perp}) }{\varepsilon_m^2 - \varepsilon_{\perp} \varepsilon_{\parallel}} }.
\label{beta_e1}
\end{equation}
 if the $c$-axis is oriented perpendicular to the interface and
\begin{equation}
\beta_e = \sqrt{\frac{\varepsilon_m \varepsilon_{\perp} (\varepsilon_m - \varepsilon_{\parallel}) }{\varepsilon_m^2 - \varepsilon_{\perp} \varepsilon_{\parallel}} }.
\label{beta_e2}
\end{equation}
if the direction of $c$-axis coincides with the SPP propagation direction. Such dependence of the dispersion on the orientation of the anisotropic crystal was presented in~\cite{Liscidini:2010}. Numerical estimations of the SPP mode "birefringence" for the two orientations of the optical axis (see Fig.~\ref{fig_geometry}b) show that the difference between the wavenumbers of the two modes increases near the surface plasmon frequency $\omega_{sp}$ at which $\beta(\omega_{sp}) = \infty$. There is a slight difference between the surface plasmon frequencies of the two modes according to Eqs.~(\ref{beta_o}), (\ref{beta_e1}) and (\ref{beta_e2}). For example, for the gold/$\mathrm{Cr_2O_3}$ interface the surface plasmon frequency corresponds to $\lambda_{sp~e} = 506$~nm and $\lambda_{sp~o} = 503.5$~nm while for the silver/$\mathrm{Cr_2O_3}$ interface $\lambda_{sp~e} = 390$~nm and $\lambda_{sp~o} = 387$~nm. For higher wavelengths ($\lambda > 900$~nm for the considered structure) the birefrigence reaches its asymptote $(\beta_e - \beta_o)/ \beta_o = \sqrt{\varepsilon_{\parallel}/\varepsilon_{\perp}} - 1$ and equals $2.6~\%$ in $\mathrm{Cr_2O_3}$.

The analysis of the SPP mode characteristics in $\mathrm{Cr_2O_3}$/metal structure was performed similar to the analysis of toroidal and axion-like plasmonic structures discussed above. Table~\ref{tabular:geometries} summarizes its results showing which changes in mode properties corresponding to the certain configurations and the crystal phases are observed. The analytical expressions for the corresponding changes are rather complicated  nevertheless Table~\ref{tabular:geometries} gives an overview of the nature of the ME impact on the SPP modes. Further we discuss several configurations in a more detailed way and perform numerical simulations using the following values for the ME coefficients: $\lambda_1 = 1.6\cdot10^{-4}$, $\lambda_2 = 1.0\cdot10^{-4}$, $\lambda_3 = -1.965\cdot10^{-4}$, $\lambda_5 = -0.82\cdot10^{-4}$ \cite{Krichevtsov:1993}.

\renewcommand{\arraystretch}{1.6} 
\begin{table}
\caption{Characteristics of SPP modes that are affected by the ME effect in $\mathrm{Cr_2O_3}$/metal structure for various configurations and phases.}
\label{tabular:geometries}
\begin{tabular}{|>{\centering\arraybackslash}p{0.25\columnwidth}|>{\centering\arraybackslash}p{0.25\columnwidth}|>{\centering\arraybackslash}p{0.2\columnwidth}|>{\centering\arraybackslash}p{0.25\columnwidth}|}
\hline
\rowcolor{Gray}[.8\tabcolsep]
\multicolumn{4}{|c|}{$c$ $\perp$ interface} \\
\hline
\rowcolor{Gray}[.8\tabcolsep]
& \multicolumn{3}{c|}{The spin-flop phase ($\mathbf{L}$ $\perp$ $c$)} \\
\hhline{|>{\arrayrulecolor{Gray}}->{\arrayrulecolor{black}}|---|}
\rowcolor{Gray}[.8\tabcolsep]
 \multirow{-2}{0.25\columnwidth}{\centering The low-field phase ($\mathbf{L}$~$\parallel$~$c$)} & \multicolumn{2}{c|}{SPP $\perp$   $\mathbf{L}$}&  SPP $\parallel$ $\mathbf{L}$\\
\hline
polarization ($\frac{E_y}{H_y}$ is real) & \multicolumn{2}{p{0.45\columnwidth}|}{\centering dispersion \\ \centering localization} & polarization ($\frac{E_y}{H_y}$ is imag.), localization in dielectric\\
\hline
\rowcolor{Gray}[.8\tabcolsep]
\multicolumn{4}{|c|}{$c$ $\parallel$ interface} \\
\hline
\rowcolor{Gray}[.8\tabcolsep] & \multicolumn{3}{c|}{The spin-flop phase ($\mathbf{L}$ $\perp$ $c$)} \\
\hhline{|>{\arrayrulecolor{Gray}}->{\arrayrulecolor{black}}|---|}
\rowcolor{Gray}[.8\tabcolsep] &  \multicolumn{2}{c|}{$\mathbf{L}$ $\perp$ interface} &  \\
\hhline{|>{\arrayrulecolor{Gray}}->{\arrayrulecolor{black}}|--|>{\arrayrulecolor{Gray}}->{\arrayrulecolor{black}}}
\rowcolor{Gray}[.8\tabcolsep] \multirow{-3}{0.25\columnwidth}{\centering The low-field phase ($\mathbf{L}$~$\parallel$~$c$), any SPP direction} & SPP $\parallel$ $c$ & SPP $\perp$ $c$  & \multirow{-2}{0.25\columnwidth}{\centering $\mathbf{L}$ $\parallel$ interface, any SPP direction}\\
\hline
polarization ($\frac{E_y}{H_y}$ is real) & polarization ($\frac{E_y}{H_y}$ is imag.)
localization in dielectric & dispersion localization & localization in dielectric\\
\hline
\end{tabular}
\end{table}

\subsection{SPP polarization in the low-field phase of $\mathrm{Cr_2O_3}$}

Let us discuss how the SPP polarization, which is purely TM in the absence of the magnetoelectricity, transforms due to the ME effect in the low-field phase of $\mathrm{Cr_2O_3}$ (which means that the antiferromagnetic vector $\mathbf{L}$ is parallel to the $c$-axis). Since this case corresponds to the axion effect described in Section IIB, one should expect change only in SPP polarization. The analysis shows that in both possible basic configurations, when the $c$-axis is oriented along the interface or perpendicular to it, the SPP mode has both nonzero TM and TE components (see Eq.~(\ref{e_spp_cr})) with different localization coefficients that are independent of magnetoelectric susceptibility tensor components. The TE/TM ratio for the "extraordinary" SPP inside $\mathrm{Cr_2O_3}$ is calculated as:
\begin{eqnarray}
\begin{split}
\frac{E_y}{H_y}  =
&\frac{\lambda_5 - \lambda_3}{\varepsilon_{\perp} - \varepsilon_{\parallel}} - \\
-&\left[(\gamma_m + \gamma_{TM})\frac{\lambda_5 - \lambda_3}{\varepsilon_{\perp} - \varepsilon_{\parallel}} + \gamma_{TM}\frac{\lambda_5}{\varepsilon_{\perp}} \right] \frac{e^{-(\gamma_{TE} - \gamma_{TM})|z|}}{\gamma_m + \gamma_{TE}}~~~
\end{split}
\label{TE-TM_af_e}
\end{eqnarray}

The analogous expression can be obtained for the "ordinary" SPP:
\begin{equation}
\begin{split}
\frac{E_y}{H_y} = &\frac{\lambda_5 - \lambda_3}{\varepsilon_{\perp} - \varepsilon_{\parallel}} - \\
-&\left[ \frac{\lambda_5 - \lambda_3}{\varepsilon_{\perp} - \varepsilon_{\parallel}} +
\frac{\lambda_3 \gamma_{TM}}{\varepsilon_{\perp} (\gamma_{TE} + \gamma_m)} \right] e^{-(\gamma_{TE} - \gamma_{TM})|z|}
\end{split}
\label{TE-TM_af_o}
\end{equation}

\begin{figure}
\centering
\subfigure[]{\includegraphics[width = 0.49\columnwidth]{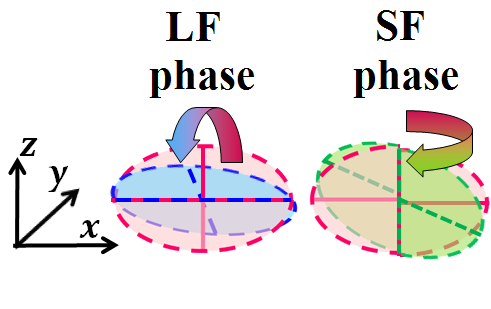}}
\subfigure[]{\includegraphics[width = 0.49\columnwidth]{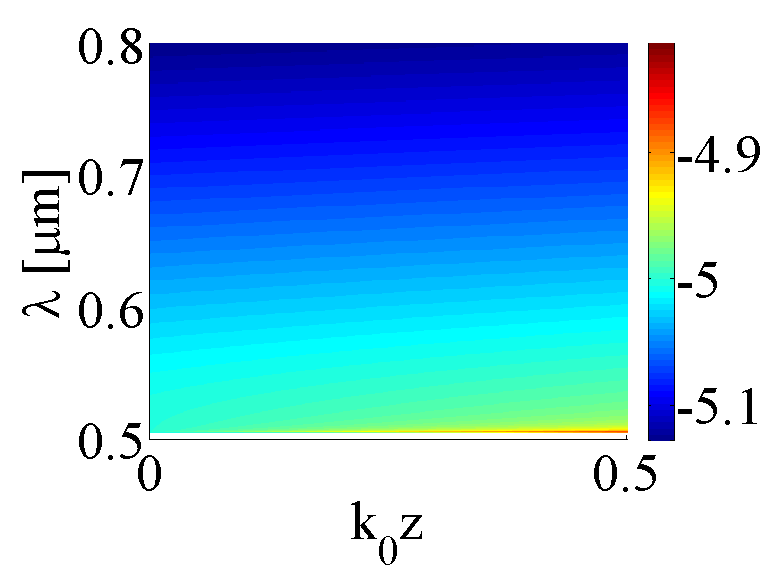}}\\
\subfigure[]{\includegraphics[width = 0.49\columnwidth]{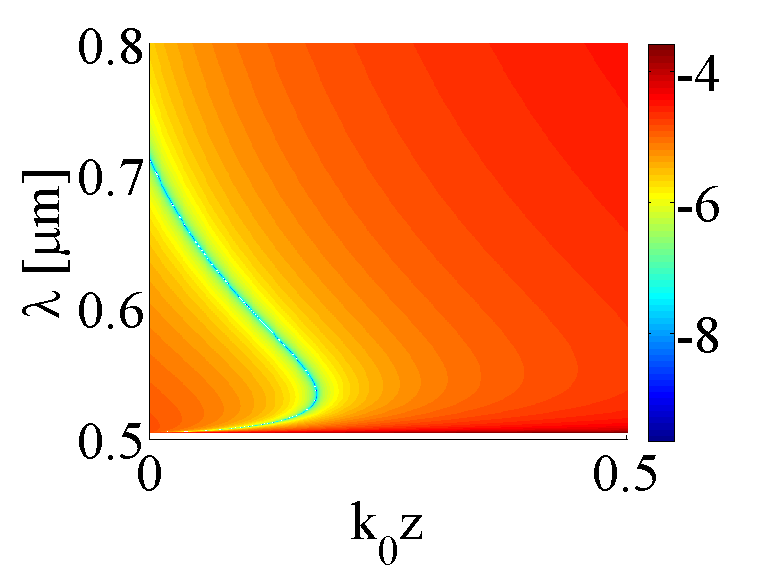}}
\subfigure[]{\includegraphics[width = 0.49\columnwidth]{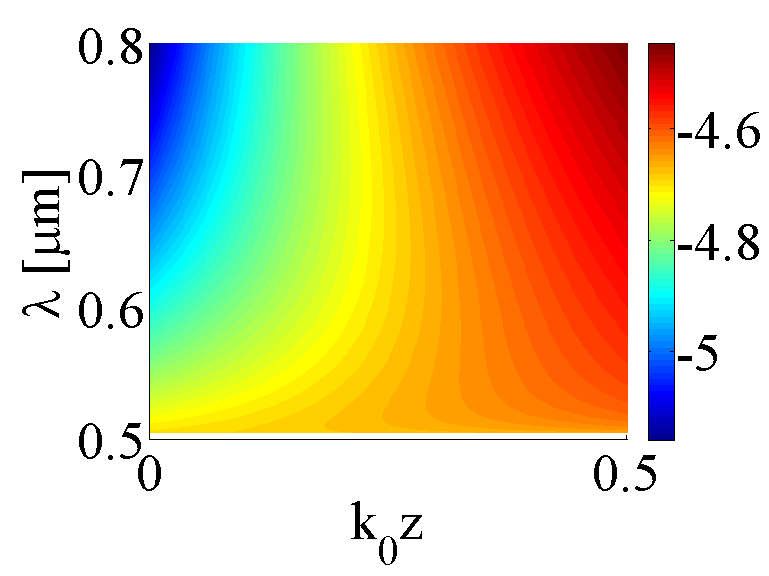}}\\

\caption{(a) The schematic representation of the corresponding tilt of the polarization ellipse for the low-field (LF) and spin-flop (SF) phases; (b-d) the dependence of the cross-section polarization of the ME SPP ($\lg|E_y/H_y|$)) on wavelength in the cases of:  (b) the spin-flop phase, extraordinary SPP; (c) the low-field phase, extraordinary SPP;  (d) the low-field phase, ordinary SPP.}
\label{figure:af_sf_polarization}
\end{figure}

In the absence of the ME effect the electric field of the SPP is elliptically polarized having both $x$- and $z$-components. According to Eqs.~(\ref{TE-TM_af_e}), (\ref{TE-TM_af_o}) $E_y$ component in the low-field state has the same phase as $H_y$ and $E_z$ components so that the polarization ellipse turns around the $x$ axis (that coincides with SPP propagation direction) with respect to the non-ME case (see Fig.~\ref{figure:af_sf_polarization}a).

It is important that the polarization of the SPP is constant during its propagation along the metal-dielectric interface. Similar effect was reported earlier for gyrotropic interfaces, in contrast to such phenomenon as Faraday rotation or optical activity-caused rotation of the bulk waves. At the same time according to Eqs.~(\ref{TE-TM_af_e}), (\ref{TE-TM_af_o}) the polarization of the SPP depends on $z$ inside $\mathrm{Cr_2O_3}$. According to the simulations presented in Fig.~\ref{figure:af_sf_polarization}c,d in both cases TE/TM ratio increases with the distance to the interface and at the distance equal to the penetration depth $k_0 z=0.5$ is significantly higher than at the interface.

Near-field polarization of the "extraordinary" SPP in the low-field phase demonstrates very peculiar behaviour. For the spectral range from 506 nm to 711 nm polarization ellipse tilt changes its sign inside the ME crystal. At the wavelength of 712 nm the polarization of the SPP is purely TM exactly at the metal-dielectric interface and is twisting with the distance inside $\mathrm{Cr_2O_3}$.

\subsection{SPP polarization in the spin-flop phase of $\mathrm{Cr_2O_3}$}
The analogous effect of the polarization ellipse tilt can be observed in the spin-flop phase of $\mathrm{Cr_2O_3}$, for example, in the configuration if the $c$-axis is oriented perpendicular to the interface while the directions of the SPP propagation and the antiferromagnetic vector $\mathbf{L}$ coincide. Comparison of Eqs.~(\ref{alpha_in_cr2o3}) and (\ref{alpha_tensor}) reveals that in this configuration the emerged toroidal moment lies in the interface plane perpendicular to the direction of the SPP propagation (see Table~\ref{tabular:idealized}).

The TE/TM ratio in this configuration can be described as follows:
\begin{equation}
\begin{split}
\frac{E_y}{H_y} = -i \frac{\lambda_2}{e^{-\gamma_{TM}|z|}} \cdot
\Big[\frac{\gamma_{TM}}{\varepsilon_{\perp} \beta} e^{-\gamma_{TM}|z|} +\\
+\frac{1}{\varepsilon_{\perp} \beta} \frac{\varepsilon_{\perp} - \gamma_{TM}\gamma_m}{\gamma_m + \gamma_{TE}}
e^{-\gamma_{TE}|z|}\Big]
\end{split}
\label{TE-TM_sf_e}
\end{equation}

In contrast to the polarization turn in the low-field phase of $\mathrm{Cr_2O_3}$, $E_y$ and $H_y$ components of the SPP have the phase shift of $\pi/2$ and $E_y$ component has the same phase as $E_x$, so that the polarization ellipse experiences a slight turn around $Oz$-axis with respect to the non-ME case (see Fig.~\ref{figure:af_sf_polarization}a)). At the same time, the SPP polarization is inhomogeneous in its cross-section analogous to the case of the low-field phase (Fig.~\ref{figure:af_sf_polarization}b).

The detailed analysis shows that in this case the ME effect also influences the localization coefficients for the two modes in $\mathrm{Cr_2O_3}$:
\begin{eqnarray}
\gamma_{TM} = \sqrt{\frac{\varepsilon_{\perp}^2 (\varepsilon_{\parallel} - \varepsilon_m)}{\varepsilon_m^2 - \varepsilon_{\parallel} \varepsilon_{\perp}}} + \lambda_1 \\
\gamma_{TE} = \sqrt{\frac{\varepsilon_{\parallel} \varepsilon_{\perp} (\varepsilon_{\perp} - \varepsilon_m) - \varepsilon_m^2 (\varepsilon_{\perp} - \varepsilon_{\parallel})}{\varepsilon_m^2 - \varepsilon_{\parallel} \varepsilon_{\perp}}} -\lambda_1
\end{eqnarray}
although the value of this ME contribution is very small since $\lambda_1\sim 10^{-4}$. Variations of the localization coefficients practically can be detected, for example, in the schemes with SPP tunneling between two metallic interfaces located close to each other.

\subsection{SPP dispersion and non-reciprocity in the spin-flop phase of $\mathrm{Cr_2O_3}$}
For definiteness we consider the following configuration: the $c$-axis is oriented perpendicular to the interface, the SPP propagation direction coincides with $Ox$ axis while the antiferromagnetic vector $\mathbf{L}$ is oriented along $Oy$ axis. At this, the toroidal moment is directed along $Ox$ axis. According to Table~\ref{tabular:idealized} in this case one may observe the non-reciprocal propagation constant variation due to the ME effect in the spin-flop phase. For the SPP propagating along the positive direction of $Ox$ axis the propagation constant has the form:
\begin{equation}
\beta = \sqrt{\frac{\varepsilon_m \varepsilon_{\parallel}(\varepsilon_m - \varepsilon_{\perp})}
{\varepsilon_m^2 - \varepsilon_{\parallel} \varepsilon_{\perp}}}
- \lambda_2 \frac{\varepsilon_m^2}{\varepsilon_m^2 - \varepsilon_{\parallel} \varepsilon_{\perp}}
\label{beta_sf}
\end{equation}

\begin{figure}
\centering
\subfigure[]{\includegraphics[width = 0.65\columnwidth]{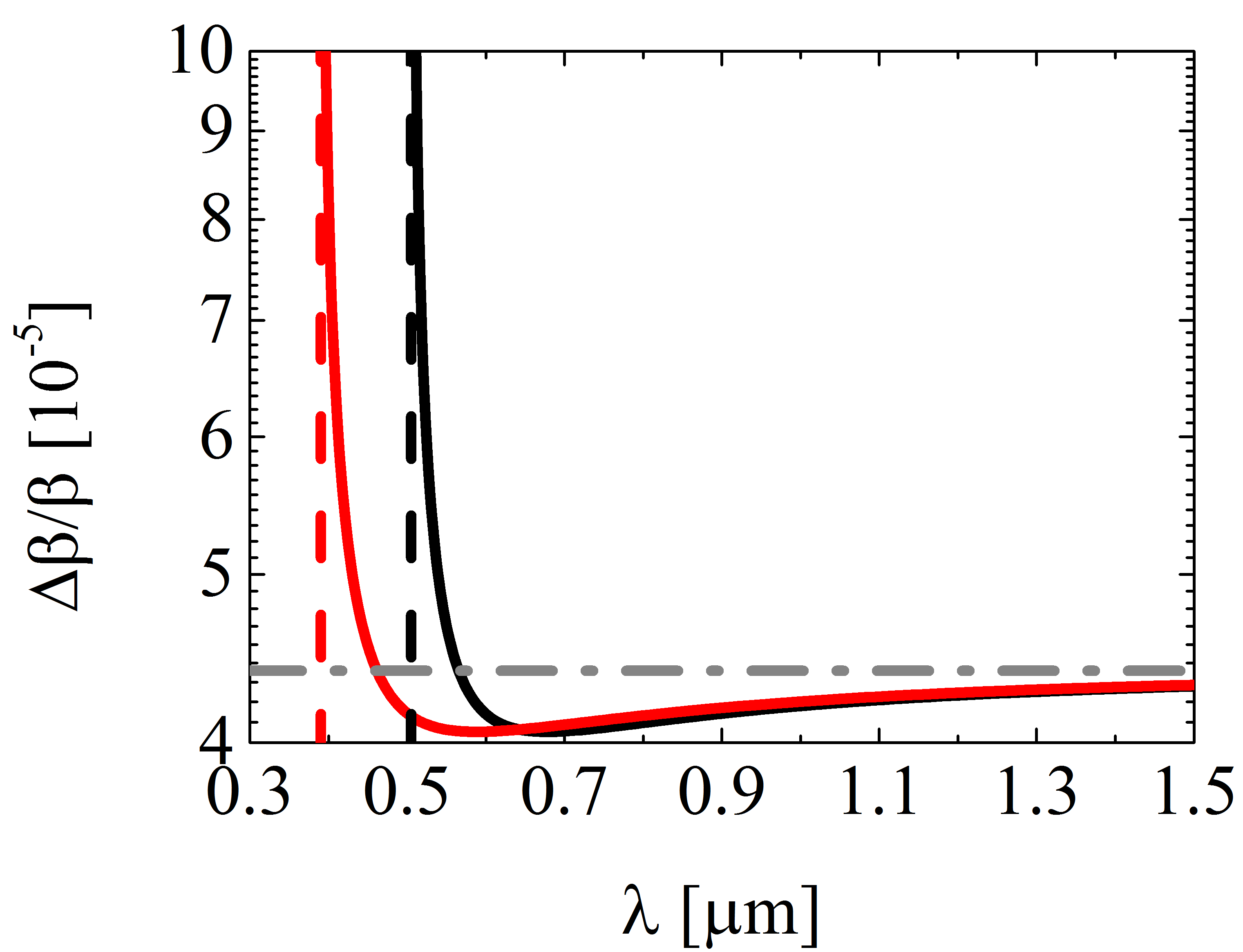} \label{figure:sf_dispersion}}
\subfigure[]{\includegraphics[width = 0.65\columnwidth]{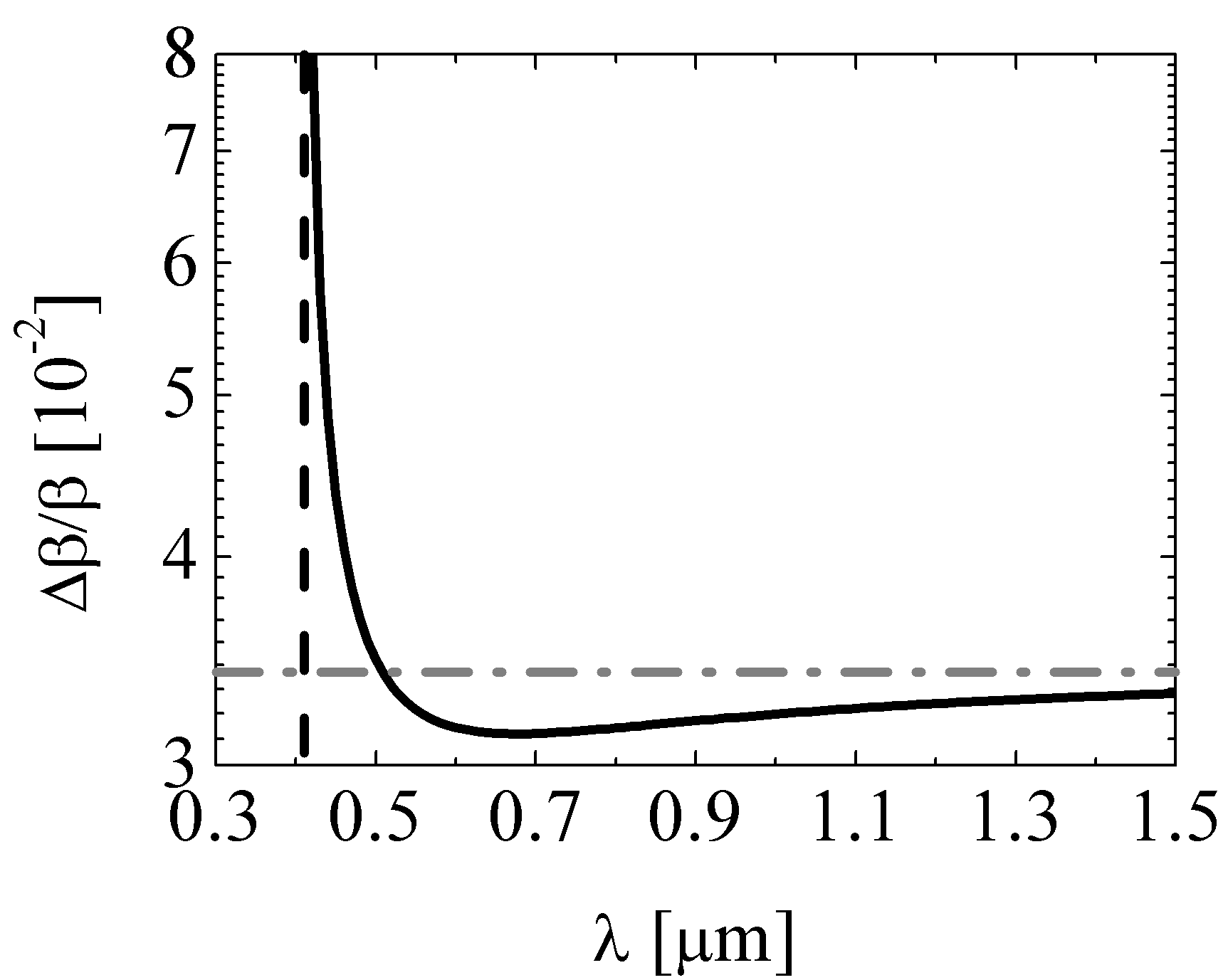} \label{figure:add}}
\caption{Dispersion variation of the SPP due to the ME effect (a) in the spin-flop phase of $\mathrm{Cr_2O_3}$, and (b) in the material with $\alpha=0.08$ (see details in the text). Red and black solid lines correspond to silver and gold interfaces, respectively. Dashed lines correspond to surface plasmon frequencies $\omega_{sp}$ and dash-dot line shows the asymptote for $\Delta \beta / \beta$.}
\end{figure}

Due to non-reciprocity the ME contribution to the propagation constant $\Delta \beta$ (the term in Eq.~(\ref{beta_sf}) that is proportional to $\lambda_2$) changes its sign if the SPP propagates in the opposite direction. The dispersion of the SPP in the considered configuration is presented in Fig.~\ref{figure:sf_dispersion}. It should be noted that ME contribution to the SPP propagation constant increases at the wavelengths near the surface plasmon frequency $\omega_{sp}$. On the contrary, $\Delta \beta$ has an asymptote $\Delta \beta / \beta = \lambda_2 / \sqrt{\varepsilon_{\parallel}}$ for higher wavelengths.

$\mathrm{Cr_2O_3}$ possesses rather low values of the magnetoelectric constants that are of about $10^{-4}$. Recently artificial structures have been constructed that exhibit magnetoelectric properties with the characteristic value of 0.1 so all the effects described in the present paper are enhanced by three orders of magnitude. An example is shown in Fig.~\ref{figure:add} where the giant non-reciprocity i.e. ME-induced SPP dispersion variation is shown for the material with $\epsilon=5.5$ and $\alpha=0.08$. These parameters correspond to the $\mathrm{BaTiO_3/BiFeO_3}$ superlattice fabricated in~\cite{Lorenz:2015}. The ME effect is about $10^{-2}$ that is three orders higher than for $\mathrm{Cr_2O_3}$ (Fig.~\ref{figure:sf_dispersion}).

At the same time, the localization coefficients, and, consequently, the SPP penetration depth experience changes both in metal and in $\mathrm{Cr_2O_3}$ as well and can be expressed for the considered configuration as:
\begin{eqnarray}
\gamma_{TM} = \sqrt{\frac{\varepsilon_{\perp}^2 (\varepsilon_{\parallel} - \varepsilon_m)}{\varepsilon_m^2 - \varepsilon_{\parallel} \varepsilon_{\perp}}} +
\Delta \beta \sqrt{-\frac{\varepsilon_m (\varepsilon_m - \varepsilon_{\perp})}{\varepsilon_{\parallel} (\varepsilon_m - \varepsilon_{\parallel}) }} \\
\gamma_{m} = \sqrt{\frac{\varepsilon_m^2 (\varepsilon_{\parallel} - \varepsilon_m)}{\varepsilon_m^2 - \varepsilon_{\parallel} \varepsilon_{\perp}}} +
\Delta \beta \sqrt{-\frac{\varepsilon_{\parallel}(\varepsilon_m - \varepsilon_{\perp})}{\varepsilon_m (\varepsilon_m - \varepsilon_{\parallel}) }}
\end{eqnarray}

Numerical estimations show that the relative variation of the localization coefficient in $\mathrm{Cr_2O_3}$ $\Delta \gamma_{TM}/\gamma_{TM}$ is about $10^{-6}$ at wavelength of $600~\text{nm}$ that is one order of magnitude higher than the corresponding variation in a metal $\Delta \gamma_{m}/\gamma_{m}$. With the increase of wavelength to about $1.5~\mu\text{m}$ $\Delta \gamma_{TM}$ becomes one order of magnitude higher while $\Delta \gamma_{m}$ experiences one-order decrease.

\subsection{Far-field control using ME SPPs}

The analysis performed above shows that ME properties of the material provide the variations of the SPP near-field properties. These variations also can be observed in the far-field as the variation of the reflection spectra of the ME/metal structure. Namely, there are two straightforward ways to get the far-field response from the SPP modulation:

1. Dispersion-based measurements. This scheme can be implemented only in spin-flop phase with SPP  $ \perp \mathbf L \perp c$ (see Table~\ref{tabular:geometries}) where ME coupling changes the propagation constant of the SPP. The shift of the surface plasmon resonance position is determined by the SPP dispersion variation due to application of the external magnetic field $\mathbf{H}$  which causes the spin re-orientation in ME crystal. This shift results in the variation of the angular reflectance spectra which can be measured as 
\begin{equation}
\delta = \frac {R(\mathbf{H})  - R(0)}{R(\mathbf{H}) +R(0)}.
\end{equation}
This idea is very similar to the well-known transverse magneto-optical Kerr effect in magnetic plasmonic structures~\cite{TMOKE_in_PlC}.

\begin{figure}
\centering
\subfigure[]{\includegraphics[width = 0.65\columnwidth]{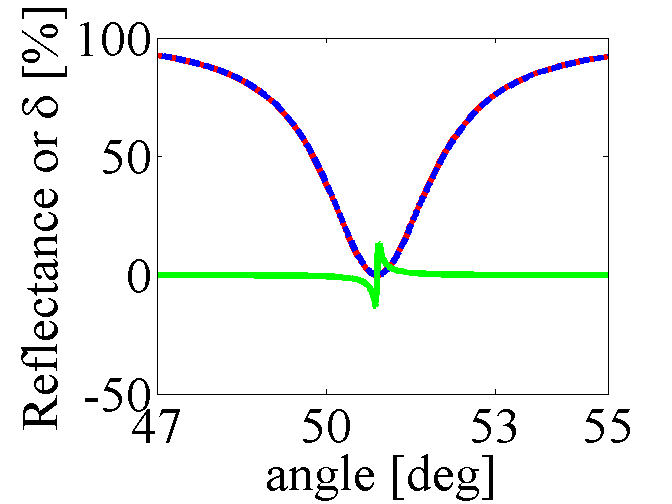} \label{figure:ff_cr}}
\subfigure[]{\includegraphics[width = 0.65\columnwidth]{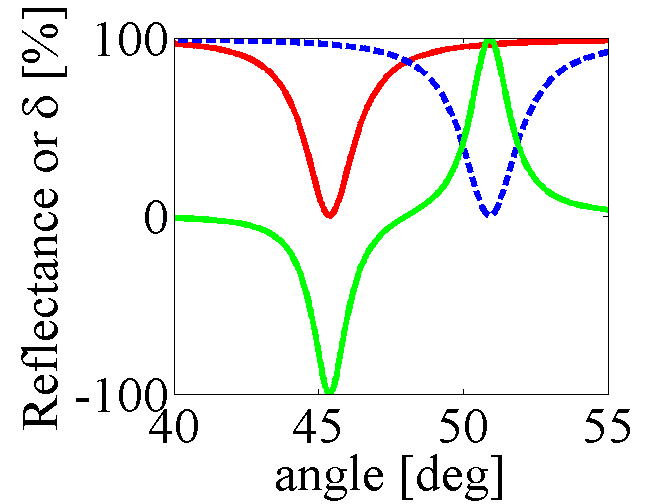} \label{figure:ff_ba}}
\caption{Reflectance angular spectra (blue and red curves for two opposite magnetic field directions) and its variation $\delta$ (green curve) due to ME switching calculated (a) in the spin-flop phase of $\mathrm{Cr_2O_3}$, and (b) in the material with $\alpha=0.08$ (see details in the text).}
\end{figure}

Let us consider ME SPPs excited at 800 nm wavelength using Kretchmann scheme via $\mathrm{GaP}$ prism at the $\mathrm{Cr_2O_3}$/gold interface in spin-flop configuration analyzed above. The variation of the reflectance via switching of the external magnetic field is shown in Fig~\ref{figure:ff_cr}. Notice that although the shift of the SPP resonance is very moderate, the magnitude of the relative reflectance modulation $\delta$ is 13$\%$ and thus can be measured using ordinary experimental techniques. For the artificial material of $\mathrm{BaTiO_3/BiFeO_3}$ superlattice (for example, reported in~\cite{Lorenz:2015}), the shift of the resonance is clearly seen in the reflectance spectra of the same prism $\mathrm{GaP}$/ME/gold structure, see Fig.~\ref{figure:ff_ba}. It can be used not only almost 100$\%$ modulation of the signal, but for the magnetic-field controlled deflection of the laser beams as well.

2. Polarization-based measurements. Since in certain cases (see Table~\ref{tabular:geometries}) a tilt of SPP polarization ellipse is observed, one can measure this tilt using either polarization or reflectance measurements. Namely, a rotation of the polarization of the reflected light can be measured (which is similar to the magnetooptical Faraday effect) or variation of the reflectance spectra due to the excitation efficiency change of the SPP with tilted polarization ellipse. Moreover, such ME SPPs can be excited using TE-polarized incident light so providing the possibility for imaging of the spin-flop effect. However the observed polarization-based far-field effects depend on the square of the ME constant and are less pronounced than the dispersion-based measurements.

\section{Conclusion}

We have performed the original thorough research of the ME influence on the SPPs. The toroidal and axion-like impact on the SPP dispersion, localization and polarization for the plasmonic modes at the interface between a dielectric with magnetoelectric properties and a metal is analyzed. The results are applied to the important special case of the metal/$\mathrm{Cr_2O_3}$ interface. We present classification of such impact for different configurations, i.e. for different relative orientations of the interface, the optical axis of $\mathrm{Cr_2O_3}$ crystal, the antiferromagnetic vector and the direction of the SPP propagation.

The most significant for practical applications is the magnetoelectric-caused variation of the SPP dispersion and polarization. Interestingly, these effects are non-reciprocal. Even the axion-like isotropic magnetoelectric effect leads to non-reciprocal tilt of the SPP polarization ellipse, whose orientation is defined by the direction of the SPP propagation. In other words, we observe locking between the SPP propagation direction and the tilt angle of the polarization ellipse, similarly to the spin-momentum locking for electronic surface states. Dispersion and polarization modulation are promising for active control of the SPP by external magnetic and electric fields. Additionally, it can be used to observe the phase transition in a $\mathrm{Cr_2O_3}$ crystal to the spin-flop phase. The changes in the SPP polarization induced by the ME effect might also be detected experimentally. In particular, the SPP can be excited by the TE-polarized incident light.

Importantly, the presented analysis is not limited to $\mathrm{Cr_2O_3}$ crystal. $\mathrm{Cr_2O_3}$ is a unique material that demonstrates switching between the two types of the magnetoelectricity and therefore is the appropriate model object for the analysis of electrodynamics. However, its significant disadvantage is rather low values of the magnetoelectric constants that are of about $10^{-4}$.  Recently artificial structures have been constructed that exhibit magnetoelectric properties with the characteristic value of 0.1 so all the effects described in the present paper are enhanced by three orders of magnitude. At this the effective manipulation of the SPP near-field and the SPP propagation can be achieved. We particularly emphasize that the described giant SPP non-reciprocity in a layered structure is an unexpected and prosperous phenomenon.

The basic principles of the magnetoelectric control of SPPs can be implemented in more complex plasmonic structures. Excitation of the ME SPPs via dielectric prisms or metallic gratings result in the appearance of the resonances in transmittance and reflectance spectra. The shift of the SPP resonance due to the spin-flop effect in the corresponding configurations will result in the variation of the transmittance and reflectance so that the spin-flop effect can be observed optically. Moreover, one can modulate the reflectance or transmittance of the structure by switching the magnetic field direction. Next, the SPPs excited by the TE-polarized incident light lead to the additional resonances in transmittance and reflectance spectra that can be of prime interest for managing of the optical response of plasmonic structures.

\acknowledgments
The work was supported by the Russian Foundation for Basic Research (projects 16-02-01065 and 16-52-00137) and the Russian Presidential Grant MK-2047.2017.2.

\bibliography{_art_bibl}

\begin{thebibliography}{59}%
\makeatletter
\providecommand \@ifxundefined [1]{%
 \@ifx{#1\undefined}
}%
\providecommand \@ifnum [1]{%
 \ifnum #1\expandafter \@firstoftwo
 \else \expandafter \@secondoftwo
 \fi
}%
\providecommand \@ifx [1]{%
 \ifx #1\expandafter \@firstoftwo
 \else \expandafter \@secondoftwo
 \fi
}%
\providecommand \natexlab [1]{#1}%
\providecommand \enquote  [1]{``#1''}%
\providecommand \bibnamefont  [1]{#1}%
\providecommand \bibfnamefont [1]{#1}%
\providecommand \citenamefont [1]{#1}%
\providecommand \href@noop [0]{\@secondoftwo}%
\providecommand \href [0]{\begingroup \@sanitize@url \@href}%
\providecommand \@href[1]{\@@startlink{#1}\@@href}%
\providecommand \@@href[1]{\endgroup#1\@@endlink}%
\providecommand \@sanitize@url [0]{\catcode `\\12\catcode `\$12\catcode
  `\&12\catcode `\#12\catcode `\^12\catcode `\_12\catcode `\%12\relax}%
\providecommand \@@startlink[1]{}%
\providecommand \@@endlink[0]{}%
\providecommand \url  [0]{\begingroup\@sanitize@url \@url }%
\providecommand \@url [1]{\endgroup\@href {#1}{\urlprefix }}%
\providecommand \urlprefix  [0]{URL }%
\providecommand \Eprint [0]{\href }%
\providecommand \doibase [0]{http://dx.doi.org/}%
\providecommand \selectlanguage [0]{\@gobble}%
\providecommand \bibinfo  [0]{\@secondoftwo}%
\providecommand \bibfield  [0]{\@secondoftwo}%
\providecommand \translation [1]{[#1]}%
\providecommand \BibitemOpen [0]{}%
\providecommand \bibitemStop [0]{}%
\providecommand \bibitemNoStop [0]{.\EOS\space}%
\providecommand \EOS [0]{\spacefactor3000\relax}%
\providecommand \BibitemShut  [1]{\csname bibitem#1\endcsname}%
\let\auto@bib@innerbib\@empty
\bibitem [{\citenamefont {MacDonald}\ \emph {et~al.}(2009)\citenamefont
  {MacDonald}, \citenamefont {Samson}, \citenamefont {Stockman},\ and\
  \citenamefont {Zheludev}}]{MacDonald:2009}%
  \BibitemOpen
  \bibfield  {author} {\bibinfo {author} {\bibfnamefont {K.~F.}\ \bibnamefont
  {MacDonald}}, \bibinfo {author} {\bibfnamefont {Z.~L.}\ \bibnamefont
  {Samson}}, \bibinfo {author} {\bibfnamefont {M.~I.}\ \bibnamefont
  {Stockman}}, \ and\ \bibinfo {author} {\bibfnamefont {N.~I.}\ \bibnamefont
  {Zheludev}},\ }\href@noop {} {\bibfield  {journal} {\bibinfo  {journal}
  {Nature Photonics}\ }\textbf {\bibinfo {volume} {3}},\ \bibinfo {pages} {55}
  (\bibinfo {year} {2009})}\BibitemShut {NoStop}%
\bibitem [{\citenamefont {Strohfeldt}\ \emph {et~al.}(2014)\citenamefont
  {Strohfeldt}, \citenamefont {Tittl}, \citenamefont {Schaferling},
  \citenamefont {Neubrech}, \citenamefont {Kreibig}, \citenamefont {Griessen},\
  and\ \citenamefont {Giessen}}]{Strohfeldt:2014}%
  \BibitemOpen
  \bibfield  {author} {\bibinfo {author} {\bibfnamefont {N.}~\bibnamefont
  {Strohfeldt}}, \bibinfo {author} {\bibfnamefont {A.}~\bibnamefont {Tittl}},
  \bibinfo {author} {\bibfnamefont {M.}~\bibnamefont {Schaferling}}, \bibinfo
  {author} {\bibfnamefont {F.}~\bibnamefont {Neubrech}}, \bibinfo {author}
  {\bibfnamefont {U.}~\bibnamefont {Kreibig}}, \bibinfo {author} {\bibfnamefont
  {R.}~\bibnamefont {Griessen}}, \ and\ \bibinfo {author} {\bibfnamefont
  {H.}~\bibnamefont {Giessen}},\ }\href@noop {} {\bibfield  {journal} {\bibinfo
   {journal} {Nano Letters}\ }\textbf {\bibinfo {volume} {14}},\ \bibinfo
  {pages} {1140} (\bibinfo {year} {2014})}\BibitemShut {NoStop}%
\bibitem [{\citenamefont {Papaioannou}\ \emph {et~al.}(2012)\citenamefont
  {Papaioannou}, \citenamefont {Kalavrouziotis}, \citenamefont {Vyrsokinos},
  \citenamefont {Weeber}, \citenamefont {Hassan}, \citenamefont {Markey},
  \citenamefont {Dereux}, \citenamefont {Kumar}, \citenamefont {Bozhevolnyi},
  \citenamefont {Baus}, \citenamefont {Tekin}, \citenamefont {Apostolopoulos},
  \citenamefont {Avramopoulos},\ and\ \citenamefont
  {Pleros}}]{Papaioannou:2012}%
  \BibitemOpen
  \bibfield  {author} {\bibinfo {author} {\bibfnamefont {S.}~\bibnamefont
  {Papaioannou}}, \bibinfo {author} {\bibfnamefont {D.}~\bibnamefont
  {Kalavrouziotis}}, \bibinfo {author} {\bibfnamefont {K.}~\bibnamefont
  {Vyrsokinos}}, \bibinfo {author} {\bibfnamefont {J.-C.}\ \bibnamefont
  {Weeber}}, \bibinfo {author} {\bibfnamefont {K.}~\bibnamefont {Hassan}},
  \bibinfo {author} {\bibfnamefont {L.}~\bibnamefont {Markey}}, \bibinfo
  {author} {\bibfnamefont {A.}~\bibnamefont {Dereux}}, \bibinfo {author}
  {\bibfnamefont {A.}~\bibnamefont {Kumar}}, \bibinfo {author} {\bibfnamefont
  {S.~I.}\ \bibnamefont {Bozhevolnyi}}, \bibinfo {author} {\bibfnamefont
  {M.}~\bibnamefont {Baus}}, \bibinfo {author} {\bibfnamefont {T.}~\bibnamefont
  {Tekin}}, \bibinfo {author} {\bibfnamefont {D.}~\bibnamefont
  {Apostolopoulos}}, \bibinfo {author} {\bibfnamefont {H.}~\bibnamefont
  {Avramopoulos}}, \ and\ \bibinfo {author} {\bibfnamefont {N.}~\bibnamefont
  {Pleros}},\ }\href@noop {} {\bibfield  {journal} {\bibinfo  {journal}
  {Scientific Reports}\ }\textbf {\bibinfo {volume} {2}},\ \bibinfo {pages}
  {652} (\bibinfo {year} {2012})}\BibitemShut {NoStop}%
\bibitem [{\citenamefont {Si}\ \emph {et~al.}(2014)\citenamefont {Si},
  \citenamefont {Zhao}, \citenamefont {Leong},\ and\ \citenamefont
  {Liu}}]{Si:2014}%
  \BibitemOpen
  \bibfield  {author} {\bibinfo {author} {\bibfnamefont {G.}~\bibnamefont
  {Si}}, \bibinfo {author} {\bibfnamefont {Y.}~\bibnamefont {Zhao}}, \bibinfo
  {author} {\bibfnamefont {E.~S.~P.}\ \bibnamefont {Leong}}, \ and\ \bibinfo
  {author} {\bibfnamefont {Y.~J.}\ \bibnamefont {Liu}},\ }\href@noop {}
  {\bibfield  {journal} {\bibinfo  {journal} {Materials}\ }\textbf {\bibinfo
  {volume} {7}},\ \bibinfo {pages} {1296} (\bibinfo {year} {2014})}\BibitemShut
  {NoStop}%
\bibitem [{\citenamefont {Cetin}\ \emph {et~al.}(2012)\citenamefont {Cetin},
  \citenamefont {Yanik}, \citenamefont {Mertir}, \citenamefont {Erramilli},
  \citenamefont {Mustecaplioglu},\ and\ \citenamefont {Altug}}]{Cetin:2012}%
  \BibitemOpen
  \bibfield  {author} {\bibinfo {author} {\bibfnamefont {A.~E.}\ \bibnamefont
  {Cetin}}, \bibinfo {author} {\bibfnamefont {A.~A.}\ \bibnamefont {Yanik}},
  \bibinfo {author} {\bibfnamefont {A.}~\bibnamefont {Mertir}}, \bibinfo
  {author} {\bibfnamefont {S.}~\bibnamefont {Erramilli}}, \bibinfo {author}
  {\bibfnamefont {O.~E.}\ \bibnamefont {Mustecaplioglu}}, \ and\ \bibinfo
  {author} {\bibfnamefont {H.}~\bibnamefont {Altug}},\ }\href@noop {}
  {\bibfield  {journal} {\bibinfo  {journal} {Applied Physics Letters}\
  }\textbf {\bibinfo {volume} {101}},\ \bibinfo {pages} {121113} (\bibinfo
  {year} {2012})}\BibitemShut {NoStop}%
\bibitem [{\citenamefont {Wang}\ \emph {et~al.}(2016)\citenamefont {Wang},
  \citenamefont {Meng}, \citenamefont {Choi}, \citenamefont {Knitter},
  \citenamefont {Kim}, \citenamefont {Cao}, \citenamefont {Shalaev},\ and\
  \citenamefont {Boltasseva}}]{shalaev}%
  \BibitemOpen
  \bibfield  {author} {\bibinfo {author} {\bibfnamefont {Z.}~\bibnamefont
  {Wang}}, \bibinfo {author} {\bibfnamefont {X.}~\bibnamefont {Meng}}, \bibinfo
  {author} {\bibfnamefont {S.~H.}\ \bibnamefont {Choi}}, \bibinfo {author}
  {\bibfnamefont {S.}~\bibnamefont {Knitter}}, \bibinfo {author} {\bibfnamefont
  {Y.~L.}\ \bibnamefont {Kim}}, \bibinfo {author} {\bibfnamefont
  {H.}~\bibnamefont {Cao}}, \bibinfo {author} {\bibfnamefont {V.~M.}\
  \bibnamefont {Shalaev}}, \ and\ \bibinfo {author} {\bibfnamefont
  {A.}~\bibnamefont {Boltasseva}},\ }\href@noop {} {\bibfield  {journal}
  {\bibinfo  {journal} {Nano letters}\ }\textbf {\bibinfo {volume} {16}},\
  \bibinfo {pages} {2471} (\bibinfo {year} {2016})}\BibitemShut {NoStop}%
\bibitem [{\citenamefont {Homola}\ \emph {et~al.}(1999)\citenamefont {Homola},
  \citenamefont {Yee},\ and\ \citenamefont {Gauglitz}}]{Homola:1999}%
  \BibitemOpen
  \bibfield  {author} {\bibinfo {author} {\bibfnamefont {J.}~\bibnamefont
  {Homola}}, \bibinfo {author} {\bibfnamefont {S.~S.}\ \bibnamefont {Yee}}, \
  and\ \bibinfo {author} {\bibfnamefont {G.}~\bibnamefont {Gauglitz}},\
  }\href@noop {} {\bibfield  {journal} {\bibinfo  {journal} {Sensors and
  Actuators B: Chemical}\ }\textbf {\bibinfo {volume} {54}},\ \bibinfo {pages}
  {3} (\bibinfo {year} {1999})}\BibitemShut {NoStop}%
\bibitem [{\citenamefont {Piliarik}\ and\ \citenamefont
  {Homola}(2009)}]{piliarik2009surface}%
  \BibitemOpen
  \bibfield  {author} {\bibinfo {author} {\bibfnamefont {M.}~\bibnamefont
  {Piliarik}}\ and\ \bibinfo {author} {\bibfnamefont {J.}~\bibnamefont
  {Homola}},\ }\href@noop {} {\bibfield  {journal} {\bibinfo  {journal} {Optics
  express}\ }\textbf {\bibinfo {volume} {17}},\ \bibinfo {pages} {16505}
  (\bibinfo {year} {2009})}\BibitemShut {NoStop}%
\bibitem [{\citenamefont {Willets}\ and\ \citenamefont
  {Van~Duyne}(2007)}]{willets2007localized}%
  \BibitemOpen
  \bibfield  {author} {\bibinfo {author} {\bibfnamefont {K.~A.}\ \bibnamefont
  {Willets}}\ and\ \bibinfo {author} {\bibfnamefont {R.~P.}\ \bibnamefont
  {Van~Duyne}},\ }\href@noop {} {\bibfield  {journal} {\bibinfo  {journal}
  {Annu. Rev. Phys. Chem.}\ }\textbf {\bibinfo {volume} {58}},\ \bibinfo
  {pages} {267} (\bibinfo {year} {2007})}\BibitemShut {NoStop}%
\bibitem [{\citenamefont {Ignatyeva}\ \emph {et~al.}(2016)\citenamefont
  {Ignatyeva}, \citenamefont {Knyazev}, \citenamefont {Kapralov}, \citenamefont
  {Dietler}, \citenamefont {Sekatskii},\ and\ \citenamefont
  {Belotelov}}]{Ignatyeva:2016}%
  \BibitemOpen
  \bibfield  {author} {\bibinfo {author} {\bibfnamefont {D.~O.}\ \bibnamefont
  {Ignatyeva}}, \bibinfo {author} {\bibfnamefont {G.~A.}\ \bibnamefont
  {Knyazev}}, \bibinfo {author} {\bibfnamefont {P.~O.}\ \bibnamefont
  {Kapralov}}, \bibinfo {author} {\bibfnamefont {G.}~\bibnamefont {Dietler}},
  \bibinfo {author} {\bibfnamefont {S.~K.}\ \bibnamefont {Sekatskii}}, \ and\
  \bibinfo {author} {\bibfnamefont {V.~I.}\ \bibnamefont {Belotelov}},\
  }\href@noop {} {\bibfield  {journal} {\bibinfo  {journal} {Scientific
  reports}\ }\textbf {\bibinfo {volume} {6}},\ \bibinfo {pages} {28077}
  (\bibinfo {year} {2016})}\BibitemShut {NoStop}%
\bibitem [{\citenamefont {Kar}\ \emph {et~al.}(2017)\citenamefont {Kar},
  \citenamefont {Goswami},\ and\ \citenamefont {Saha}}]{kar2017long}%
  \BibitemOpen
  \bibfield  {author} {\bibinfo {author} {\bibfnamefont {A.}~\bibnamefont
  {Kar}}, \bibinfo {author} {\bibfnamefont {N.}~\bibnamefont {Goswami}}, \ and\
  \bibinfo {author} {\bibfnamefont {A.}~\bibnamefont {Saha}},\ }\href@noop {}
  {\bibfield  {journal} {\bibinfo  {journal} {Applied Physics B}\ }\textbf
  {\bibinfo {volume} {123}},\ \bibinfo {pages} {1} (\bibinfo {year}
  {2017})}\BibitemShut {NoStop}%
\bibitem [{\citenamefont {Chen}\ \emph {et~al.}(2011)\citenamefont {Chen},
  \citenamefont {Li}, \citenamefont {Yue},\ and\ \citenamefont
  {Gong}}]{chen2011highly}%
  \BibitemOpen
  \bibfield  {author} {\bibinfo {author} {\bibfnamefont {J.}~\bibnamefont
  {Chen}}, \bibinfo {author} {\bibfnamefont {Z.}~\bibnamefont {Li}}, \bibinfo
  {author} {\bibfnamefont {S.}~\bibnamefont {Yue}}, \ and\ \bibinfo {author}
  {\bibfnamefont {Q.}~\bibnamefont {Gong}},\ }\href@noop {} {\bibfield
  {journal} {\bibinfo  {journal} {Nano letters}\ }\textbf {\bibinfo {volume}
  {11}},\ \bibinfo {pages} {2933} (\bibinfo {year} {2011})}\BibitemShut
  {NoStop}%
\bibitem [{\citenamefont {Lu}\ \emph {et~al.}(2011)\citenamefont {Lu},
  \citenamefont {Liu}, \citenamefont {Wang}, \citenamefont {Gong},\ and\
  \citenamefont {Mao}}]{lu2011ultrafast}%
  \BibitemOpen
  \bibfield  {author} {\bibinfo {author} {\bibfnamefont {H.}~\bibnamefont
  {Lu}}, \bibinfo {author} {\bibfnamefont {X.}~\bibnamefont {Liu}}, \bibinfo
  {author} {\bibfnamefont {L.}~\bibnamefont {Wang}}, \bibinfo {author}
  {\bibfnamefont {Y.}~\bibnamefont {Gong}}, \ and\ \bibinfo {author}
  {\bibfnamefont {D.}~\bibnamefont {Mao}},\ }\href@noop {} {\bibfield
  {journal} {\bibinfo  {journal} {Optics express}\ }\textbf {\bibinfo {volume}
  {19}},\ \bibinfo {pages} {2910} (\bibinfo {year} {2011})}\BibitemShut
  {NoStop}%
\bibitem [{\citenamefont {Rotenberg}\ \emph {et~al.}(2010)\citenamefont
  {Rotenberg}, \citenamefont {Betz},\ and\ \citenamefont {van
  Driel}}]{rotenberg2010ultrafast}%
  \BibitemOpen
  \bibfield  {author} {\bibinfo {author} {\bibfnamefont {N.}~\bibnamefont
  {Rotenberg}}, \bibinfo {author} {\bibfnamefont {M.}~\bibnamefont {Betz}}, \
  and\ \bibinfo {author} {\bibfnamefont {H.~M.}\ \bibnamefont {van Driel}},\
  }\href@noop {} {\bibfield  {journal} {\bibinfo  {journal} {Physical review
  letters}\ }\textbf {\bibinfo {volume} {105}},\ \bibinfo {pages} {017402}
  (\bibinfo {year} {2010})}\BibitemShut {NoStop}%
\bibitem [{\citenamefont {Kreilkamp}\ \emph {et~al.}(2016)\citenamefont
  {Kreilkamp}, \citenamefont {Akimov}, \citenamefont {Belotelov}, \citenamefont
  {Glavin}, \citenamefont {Litvin}, \citenamefont {Rudzinski}, \citenamefont
  {Kahl}, \citenamefont {Jede}, \citenamefont {Wiater}, \citenamefont
  {Wojtowicz} \emph {et~al.}}]{kreilkamp2016terahertz}%
  \BibitemOpen
  \bibfield  {author} {\bibinfo {author} {\bibfnamefont {L.~E.}\ \bibnamefont
  {Kreilkamp}}, \bibinfo {author} {\bibfnamefont {I.~A.}\ \bibnamefont
  {Akimov}}, \bibinfo {author} {\bibfnamefont {V.~I.}\ \bibnamefont
  {Belotelov}}, \bibinfo {author} {\bibfnamefont {B.~A.}\ \bibnamefont
  {Glavin}}, \bibinfo {author} {\bibfnamefont {L.~V.}\ \bibnamefont {Litvin}},
  \bibinfo {author} {\bibfnamefont {A.}~\bibnamefont {Rudzinski}}, \bibinfo
  {author} {\bibfnamefont {M.}~\bibnamefont {Kahl}}, \bibinfo {author}
  {\bibfnamefont {R.}~\bibnamefont {Jede}}, \bibinfo {author} {\bibfnamefont
  {M.}~\bibnamefont {Wiater}}, \bibinfo {author} {\bibfnamefont
  {T.}~\bibnamefont {Wojtowicz}},  \emph {et~al.},\ }\href@noop {} {\bibfield
  {journal} {\bibinfo  {journal} {Physical Review B}\ }\textbf {\bibinfo
  {volume} {93}},\ \bibinfo {pages} {125404} (\bibinfo {year}
  {2016})}\BibitemShut {NoStop}%
\bibitem [{\citenamefont {Ignatyeva}\ and\ \citenamefont
  {Sukhorukov}(2012)}]{ignatyeva:2012:APA}%
  \BibitemOpen
  \bibfield  {author} {\bibinfo {author} {\bibfnamefont {D.~O.}\ \bibnamefont
  {Ignatyeva}}\ and\ \bibinfo {author} {\bibfnamefont {A.~P.}\ \bibnamefont
  {Sukhorukov}},\ }\href@noop {} {\bibfield  {journal} {\bibinfo  {journal}
  {Applied Physics A}\ }\textbf {\bibinfo {volume} {109}},\ \bibinfo {pages}
  {813} (\bibinfo {year} {2012})}\BibitemShut {NoStop}%
\bibitem [{\citenamefont {Ignatyeva}\ and\ \citenamefont
  {Sukhorukov}(2014)}]{ignatyeva2014femtosecond}%
  \BibitemOpen
  \bibfield  {author} {\bibinfo {author} {\bibfnamefont {D.~O.}\ \bibnamefont
  {Ignatyeva}}\ and\ \bibinfo {author} {\bibfnamefont {A.~P.}\ \bibnamefont
  {Sukhorukov}},\ }\href@noop {} {\bibfield  {journal} {\bibinfo  {journal}
  {Physical Review A}\ }\textbf {\bibinfo {volume} {89}},\ \bibinfo {pages}
  {013850} (\bibinfo {year} {2014})}\BibitemShut {NoStop}%
\bibitem [{\citenamefont {Temnov}\ \emph {et~al.}(2010)\citenamefont {Temnov},
  \citenamefont {Armelles}, \citenamefont {Woggon}, \citenamefont {Guzatov},
  \citenamefont {Cebollada}, \citenamefont {Garcia-Martin}, \citenamefont
  {Garcia-Martin}, \citenamefont {Thomay}, \citenamefont {Leitenstorfer},\ and\
  \citenamefont {Bratschitsch}}]{Temnov:2010}%
  \BibitemOpen
  \bibfield  {author} {\bibinfo {author} {\bibfnamefont {V.~V.}\ \bibnamefont
  {Temnov}}, \bibinfo {author} {\bibfnamefont {G.}~\bibnamefont {Armelles}},
  \bibinfo {author} {\bibfnamefont {U.}~\bibnamefont {Woggon}}, \bibinfo
  {author} {\bibfnamefont {D.}~\bibnamefont {Guzatov}}, \bibinfo {author}
  {\bibfnamefont {A.}~\bibnamefont {Cebollada}}, \bibinfo {author}
  {\bibfnamefont {A.}~\bibnamefont {Garcia-Martin}}, \bibinfo {author}
  {\bibfnamefont {J.-M.}\ \bibnamefont {Garcia-Martin}}, \bibinfo {author}
  {\bibfnamefont {T.}~\bibnamefont {Thomay}}, \bibinfo {author} {\bibfnamefont
  {A.}~\bibnamefont {Leitenstorfer}}, \ and\ \bibinfo {author} {\bibfnamefont
  {R.}~\bibnamefont {Bratschitsch}},\ }\href@noop {} {\bibfield  {journal}
  {\bibinfo  {journal} {Nature Photonics}\ }\textbf {\bibinfo {volume} {4}},\
  \bibinfo {pages} {107} (\bibinfo {year} {2010})}\BibitemShut {NoStop}%
\bibitem [{\citenamefont {Shcherbakov}\ \emph {et~al.}(2014)\citenamefont
  {Shcherbakov}, \citenamefont {Vabishchevich}, \citenamefont {Frolov},
  \citenamefont {Dolgova},\ and\ \citenamefont
  {Fedyanin}}]{shcherbakov2014femtosecond}%
  \BibitemOpen
  \bibfield  {author} {\bibinfo {author} {\bibfnamefont {M.}~\bibnamefont
  {Shcherbakov}}, \bibinfo {author} {\bibfnamefont {P.}~\bibnamefont
  {Vabishchevich}}, \bibinfo {author} {\bibfnamefont {A.~Y.}\ \bibnamefont
  {Frolov}}, \bibinfo {author} {\bibfnamefont {T.}~\bibnamefont {Dolgova}}, \
  and\ \bibinfo {author} {\bibfnamefont {A.}~\bibnamefont {Fedyanin}},\
  }\href@noop {} {\bibfield  {journal} {\bibinfo  {journal} {Physical Review
  B}\ }\textbf {\bibinfo {volume} {90}},\ \bibinfo {pages} {201405} (\bibinfo
  {year} {2014})}\BibitemShut {NoStop}%
\bibitem [{\citenamefont {Kalish}\ \emph {et~al.}(2014)\citenamefont {Kalish},
  \citenamefont {Ignatyeva}, \citenamefont {Bayer}, \citenamefont {Belotelov},
  \citenamefont {Kreilkamp},\ and\ \citenamefont {Sukhorukov}}]{Kalish:2014}%
  \BibitemOpen
  \bibfield  {author} {\bibinfo {author} {\bibfnamefont {A.}~\bibnamefont
  {Kalish}}, \bibinfo {author} {\bibfnamefont {D.}~\bibnamefont {Ignatyeva}},
  \bibinfo {author} {\bibfnamefont {M.}~\bibnamefont {Bayer}}, \bibinfo
  {author} {\bibfnamefont {V.}~\bibnamefont {Belotelov}}, \bibinfo {author}
  {\bibfnamefont {L.}~\bibnamefont {Kreilkamp}}, \ and\ \bibinfo {author}
  {\bibfnamefont {A.}~\bibnamefont {Sukhorukov}},\ }\href@noop {} {\bibfield
  {journal} {\bibinfo  {journal} {Laser Physics}\ }\textbf {\bibinfo {volume}
  {24}},\ \bibinfo {pages} {094006} (\bibinfo {year} {2014})}\BibitemShut
  {NoStop}%
\bibitem [{\citenamefont {Belotelov}\ \emph {et~al.}(2014)\citenamefont
  {Belotelov}, \citenamefont {Kreilkamp}, \citenamefont {Kalish}, \citenamefont
  {Akimov}, \citenamefont {Bykov}, \citenamefont {Kasture}, \citenamefont
  {Yallapragada}, \citenamefont {Gopal}, \citenamefont {Grishin}, \citenamefont
  {Khartsev}, \citenamefont {Nur-E-Alam}, \citenamefont {Vasiliev},
  \citenamefont {Doskolovich}, \citenamefont {Yakovlev}, \citenamefont
  {Alameh}, \citenamefont {Zvezdin},\ and\ \citenamefont
  {Bayer}}]{Belotelov:2014}%
  \BibitemOpen
  \bibfield  {author} {\bibinfo {author} {\bibfnamefont {V.~I.}\ \bibnamefont
  {Belotelov}}, \bibinfo {author} {\bibfnamefont {L.~E.}\ \bibnamefont
  {Kreilkamp}}, \bibinfo {author} {\bibfnamefont {A.~N.}\ \bibnamefont
  {Kalish}}, \bibinfo {author} {\bibfnamefont {I.~A.}\ \bibnamefont {Akimov}},
  \bibinfo {author} {\bibfnamefont {D.~A.}\ \bibnamefont {Bykov}}, \bibinfo
  {author} {\bibfnamefont {S.}~\bibnamefont {Kasture}}, \bibinfo {author}
  {\bibfnamefont {V.~J.}\ \bibnamefont {Yallapragada}}, \bibinfo {author}
  {\bibfnamefont {A.~V.}\ \bibnamefont {Gopal}}, \bibinfo {author}
  {\bibfnamefont {A.~M.}\ \bibnamefont {Grishin}}, \bibinfo {author}
  {\bibfnamefont {S.~I.}\ \bibnamefont {Khartsev}}, \bibinfo {author}
  {\bibfnamefont {M.}~\bibnamefont {Nur-E-Alam}}, \bibinfo {author}
  {\bibfnamefont {M.}~\bibnamefont {Vasiliev}}, \bibinfo {author}
  {\bibfnamefont {L.~L.}\ \bibnamefont {Doskolovich}}, \bibinfo {author}
  {\bibfnamefont {D.~R.}\ \bibnamefont {Yakovlev}}, \bibinfo {author}
  {\bibfnamefont {K.}~\bibnamefont {Alameh}}, \bibinfo {author} {\bibfnamefont
  {A.~K.}\ \bibnamefont {Zvezdin}}, \ and\ \bibinfo {author} {\bibfnamefont
  {M.}~\bibnamefont {Bayer}},\ }\href@noop {} {\bibfield  {journal} {\bibinfo
  {journal} {Phys. Rev. B}\ }\textbf {\bibinfo {volume} {89}},\ \bibinfo
  {pages} {045118} (\bibinfo {year} {2014})}\BibitemShut {NoStop}%
\bibitem [{\citenamefont {Belotelov}\ \emph {et~al.}(2009)\citenamefont
  {Belotelov}, \citenamefont {Bykov}, \citenamefont {Doskolovich},
  \citenamefont {Kalish}, \citenamefont {Kotov},\ and\ \citenamefont
  {Zvezdin}}]{belotelov2009giant}%
  \BibitemOpen
  \bibfield  {author} {\bibinfo {author} {\bibfnamefont {V.}~\bibnamefont
  {Belotelov}}, \bibinfo {author} {\bibfnamefont {D.}~\bibnamefont {Bykov}},
  \bibinfo {author} {\bibfnamefont {L.}~\bibnamefont {Doskolovich}}, \bibinfo
  {author} {\bibfnamefont {A.}~\bibnamefont {Kalish}}, \bibinfo {author}
  {\bibfnamefont {V.}~\bibnamefont {Kotov}}, \ and\ \bibinfo {author}
  {\bibfnamefont {A.}~\bibnamefont {Zvezdin}},\ }\href@noop {} {\bibfield
  {journal} {\bibinfo  {journal} {Optics letters}\ }\textbf {\bibinfo {volume}
  {34}},\ \bibinfo {pages} {398} (\bibinfo {year} {2009})}\BibitemShut
  {NoStop}%
\bibitem [{\citenamefont {Ignatyeva}\ \emph {et~al.}(2012)\citenamefont
  {Ignatyeva}, \citenamefont {Kalish}, \citenamefont {Levkina},\ and\
  \citenamefont {Sukhorukov}}]{Ignatyeva:2012:PRA}%
  \BibitemOpen
  \bibfield  {author} {\bibinfo {author} {\bibfnamefont {D.~O.}\ \bibnamefont
  {Ignatyeva}}, \bibinfo {author} {\bibfnamefont {A.~N.}\ \bibnamefont
  {Kalish}}, \bibinfo {author} {\bibfnamefont {G.~Y.}\ \bibnamefont {Levkina}},
  \ and\ \bibinfo {author} {\bibfnamefont {A.~P.}\ \bibnamefont {Sukhorukov}},\
  }\href@noop {} {\bibfield  {journal} {\bibinfo  {journal} {Physical Review
  A}\ }\textbf {\bibinfo {volume} {85}},\ \bibinfo {pages} {043804} (\bibinfo
  {year} {2012})}\BibitemShut {NoStop}%
\bibitem [{\citenamefont {Sukhorukov}\ \emph {et~al.}(2011)\citenamefont
  {Sukhorukov}, \citenamefont {Ignatyeva},\ and\ \citenamefont
  {Kalish}}]{Ignatyeva:2011:THZ}%
  \BibitemOpen
  \bibfield  {author} {\bibinfo {author} {\bibfnamefont {A.~P.}\ \bibnamefont
  {Sukhorukov}}, \bibinfo {author} {\bibfnamefont {D.~O.}\ \bibnamefont
  {Ignatyeva}}, \ and\ \bibinfo {author} {\bibfnamefont {A.~N.}\ \bibnamefont
  {Kalish}},\ }\href@noop {} {\bibfield  {journal} {\bibinfo  {journal}
  {Journal of Infrared, Millimeter, and Terahertz Waves}\ }\textbf {\bibinfo
  {volume} {32}},\ \bibinfo {pages} {1223} (\bibinfo {year}
  {2011})}\BibitemShut {NoStop}%
\bibitem [{\citenamefont {Mi}\ and\ \citenamefont {Van}(2014)}]{Guangcan:2014}%
  \BibitemOpen
  \bibfield  {author} {\bibinfo {author} {\bibfnamefont {G.}~\bibnamefont
  {Mi}}\ and\ \bibinfo {author} {\bibfnamefont {V.}~\bibnamefont {Van}},\
  }\href@noop {} {\bibfield  {journal} {\bibinfo  {journal} {Optics Letters}\
  }\textbf {\bibinfo {volume} {39}},\ \bibinfo {pages} {2028} (\bibinfo {year}
  {2014})}\BibitemShut {NoStop}%
\bibitem [{\citenamefont {Ignatyeva}\ \emph {et~al.}(2017)\citenamefont
  {Ignatyeva}, \citenamefont {Kalish}, \citenamefont {Belotelov},\ and\
  \citenamefont {Zvezdin}}]{Ignatyeva:2017}%
  \BibitemOpen
  \bibfield  {author} {\bibinfo {author} {\bibfnamefont {D.}~\bibnamefont
  {Ignatyeva}}, \bibinfo {author} {\bibfnamefont {A.}~\bibnamefont {Kalish}},
  \bibinfo {author} {\bibfnamefont {V.}~\bibnamefont {Belotelov}}, \ and\
  \bibinfo {author} {\bibfnamefont {A.}~\bibnamefont {Zvezdin}},\ }\href@noop
  {} {\bibfield  {journal} {\bibinfo  {journal} {Physics of Wave Phenomena}\
  }\textbf {\bibinfo {volume} {25}},\ \bibinfo {pages} {119} (\bibinfo {year}
  {2017})}\BibitemShut {NoStop}%
\bibitem [{\citenamefont {Tarkhanyan}\ and\ \citenamefont
  {Niarchos}(2011)}]{tarkhanyan2011nonradiative}%
  \BibitemOpen
  \bibfield  {author} {\bibinfo {author} {\bibfnamefont {R.~H.}\ \bibnamefont
  {Tarkhanyan}}\ and\ \bibinfo {author} {\bibfnamefont {D.~G.}\ \bibnamefont
  {Niarchos}},\ }\href@noop {} {\bibfield  {journal} {\bibinfo  {journal}
  {Phys. Status Solidi B}\ }\textbf {\bibinfo {volume} {248}},\ \bibinfo
  {pages} {1499} (\bibinfo {year} {2011})}\BibitemShut {NoStop}%
\bibitem [{\citenamefont {Galynsky}\ \emph {et~al.}(2004)\citenamefont
  {Galynsky}, \citenamefont {Furs},\ and\ \citenamefont
  {Barkovsky}}]{galynsky2004integral}%
  \BibitemOpen
  \bibfield  {author} {\bibinfo {author} {\bibfnamefont {V.~M.}\ \bibnamefont
  {Galynsky}}, \bibinfo {author} {\bibfnamefont {A.}~\bibnamefont {Furs}}, \
  and\ \bibinfo {author} {\bibfnamefont {L.}~\bibnamefont {Barkovsky}},\
  }\href@noop {} {\bibfield  {journal} {\bibinfo  {journal} {Journal of Physics
  A: Mathematical and General}\ }\textbf {\bibinfo {volume} {37}},\ \bibinfo
  {pages} {5083} (\bibinfo {year} {2004})}\BibitemShut {NoStop}%
\bibitem [{\citenamefont {Shuvaev}\ \emph {et~al.}(2011)\citenamefont
  {Shuvaev}, \citenamefont {Engelbrecht}, \citenamefont {Wunderlich},
  \citenamefont {Schneider},\ and\ \citenamefont {Pimenov}}]{Shuvaev:2011}%
  \BibitemOpen
  \bibfield  {author} {\bibinfo {author} {\bibfnamefont {A.~M.}\ \bibnamefont
  {Shuvaev}}, \bibinfo {author} {\bibfnamefont {S.}~\bibnamefont
  {Engelbrecht}}, \bibinfo {author} {\bibfnamefont {M.}~\bibnamefont
  {Wunderlich}}, \bibinfo {author} {\bibfnamefont {A.}~\bibnamefont
  {Schneider}}, \ and\ \bibinfo {author} {\bibfnamefont {A.}~\bibnamefont
  {Pimenov}},\ }\href@noop {} {\bibfield  {journal} {\bibinfo  {journal} {The
  European Physical Journal B}\ }\textbf {\bibinfo {volume} {79}},\ \bibinfo
  {pages} {163} (\bibinfo {year} {2011})}\BibitemShut {NoStop}%
\bibitem [{\citenamefont {Huang}\ \emph {et~al.}(2012)\citenamefont {Huang},
  \citenamefont {Chen}, \citenamefont {Wu}, \citenamefont {Fedotov},
  \citenamefont {Savinov}, \citenamefont {Ho}, \citenamefont {Chau},
  \citenamefont {Zheludev},\ and\ \citenamefont {Tsai}}]{Zheludev:2012}%
  \BibitemOpen
  \bibfield  {author} {\bibinfo {author} {\bibfnamefont {Y.-W.}\ \bibnamefont
  {Huang}}, \bibinfo {author} {\bibfnamefont {W.~T.}\ \bibnamefont {Chen}},
  \bibinfo {author} {\bibfnamefont {P.~C.}\ \bibnamefont {Wu}}, \bibinfo
  {author} {\bibfnamefont {V.}~\bibnamefont {Fedotov}}, \bibinfo {author}
  {\bibfnamefont {V.}~\bibnamefont {Savinov}}, \bibinfo {author} {\bibfnamefont
  {Y.~Z.}\ \bibnamefont {Ho}}, \bibinfo {author} {\bibfnamefont {Y.-F.}\
  \bibnamefont {Chau}}, \bibinfo {author} {\bibfnamefont {N.~I.}\ \bibnamefont
  {Zheludev}}, \ and\ \bibinfo {author} {\bibfnamefont {D.~P.}\ \bibnamefont
  {Tsai}},\ }\href@noop {} {\bibfield  {journal} {\bibinfo  {journal} {Opt.
  Express}\ }\textbf {\bibinfo {volume} {20}},\ \bibinfo {pages} {1760}
  (\bibinfo {year} {2012})}\BibitemShut {NoStop}%
\bibitem [{\citenamefont {Popkov}\ \emph {et~al.}(2016)\citenamefont {Popkov},
  \citenamefont {Davydova}, \citenamefont {Zvezdin}, \citenamefont
  {Solov'yov},\ and\ \citenamefont {Zvezdin}}]{popkov2016origin}%
  \BibitemOpen
  \bibfield  {author} {\bibinfo {author} {\bibfnamefont {A.}~\bibnamefont
  {Popkov}}, \bibinfo {author} {\bibfnamefont {M.}~\bibnamefont {Davydova}},
  \bibinfo {author} {\bibfnamefont {K.}~\bibnamefont {Zvezdin}}, \bibinfo
  {author} {\bibfnamefont {S.}~\bibnamefont {Solov'yov}}, \ and\ \bibinfo
  {author} {\bibfnamefont {A.}~\bibnamefont {Zvezdin}},\ }\href@noop {}
  {\bibfield  {journal} {\bibinfo  {journal} {Physical Review B}\ }\textbf
  {\bibinfo {volume} {93}},\ \bibinfo {pages} {094435} (\bibinfo {year}
  {2016})}\BibitemShut {NoStop}%
\bibitem [{\citenamefont {Bichurin}\ and\ \citenamefont
  {Viehland}(2011)}]{bichurin2011magnetoelectricity}%
  \BibitemOpen
  \bibfield  {author} {\bibinfo {author} {\bibfnamefont {M.}~\bibnamefont
  {Bichurin}}\ and\ \bibinfo {author} {\bibfnamefont {D.}~\bibnamefont
  {Viehland}},\ }\href@noop {} {\emph {\bibinfo {title} {Magnetoelectricity in
  composites}}}\ (\bibinfo  {publisher} {CRC Press},\ \bibinfo {year}
  {2011})\BibitemShut {NoStop}%
\bibitem [{\citenamefont {Sreenivasulu}\ \emph {et~al.}(2012)\citenamefont
  {Sreenivasulu}, \citenamefont {Petrov}, \citenamefont {Fetisov},
  \citenamefont {Fetisov},\ and\ \citenamefont
  {Srinivasan}}]{sreenivasulu2012magnetoelectric}%
  \BibitemOpen
  \bibfield  {author} {\bibinfo {author} {\bibfnamefont {G.}~\bibnamefont
  {Sreenivasulu}}, \bibinfo {author} {\bibfnamefont {V.}~\bibnamefont
  {Petrov}}, \bibinfo {author} {\bibfnamefont {L.}~\bibnamefont {Fetisov}},
  \bibinfo {author} {\bibfnamefont {Y.}~\bibnamefont {Fetisov}}, \ and\
  \bibinfo {author} {\bibfnamefont {G.}~\bibnamefont {Srinivasan}},\
  }\href@noop {} {\bibfield  {journal} {\bibinfo  {journal} {Physical Review
  B}\ }\textbf {\bibinfo {volume} {86}},\ \bibinfo {pages} {214405} (\bibinfo
  {year} {2012})}\BibitemShut {NoStop}%
\bibitem [{\citenamefont {Lorenz}\ \emph {et~al.}(2015)\citenamefont {Lorenz},
  \citenamefont {Wagner}, \citenamefont {Lazenka}, \citenamefont
  {Schwinkendorf}, \citenamefont {Modarresi}, \citenamefont {Van~Bael},
  \citenamefont {Vantomme}, \citenamefont {Temst}, \citenamefont {Oeckler},\
  and\ \citenamefont {Grundmann}}]{Lorenz:2015}%
  \BibitemOpen
  \bibfield  {author} {\bibinfo {author} {\bibfnamefont {M.}~\bibnamefont
  {Lorenz}}, \bibinfo {author} {\bibfnamefont {G.}~\bibnamefont {Wagner}},
  \bibinfo {author} {\bibfnamefont {V.}~\bibnamefont {Lazenka}}, \bibinfo
  {author} {\bibfnamefont {P.}~\bibnamefont {Schwinkendorf}}, \bibinfo {author}
  {\bibfnamefont {H.}~\bibnamefont {Modarresi}}, \bibinfo {author}
  {\bibfnamefont {M.~J.}\ \bibnamefont {Van~Bael}}, \bibinfo {author}
  {\bibfnamefont {A.}~\bibnamefont {Vantomme}}, \bibinfo {author}
  {\bibfnamefont {K.}~\bibnamefont {Temst}}, \bibinfo {author} {\bibfnamefont
  {O.}~\bibnamefont {Oeckler}}, \ and\ \bibinfo {author} {\bibfnamefont
  {M.}~\bibnamefont {Grundmann}},\ }\href@noop {} {\bibfield  {journal}
  {\bibinfo  {journal} {Applied Physics Letters}\ }\textbf {\bibinfo {volume}
  {106}},\ \bibinfo {pages} {012905} (\bibinfo {year} {2015})}\BibitemShut
  {NoStop}%
\bibitem [{\citenamefont {Eerenstein}\ \emph {et~al.}(2006)\citenamefont
  {Eerenstein}, \citenamefont {Mathur},\ and\ \citenamefont
  {Scott}}]{Eerenstein:2006}%
  \BibitemOpen
  \bibfield  {author} {\bibinfo {author} {\bibfnamefont {W.}~\bibnamefont
  {Eerenstein}}, \bibinfo {author} {\bibfnamefont {N.}~\bibnamefont {Mathur}},
  \ and\ \bibinfo {author} {\bibfnamefont {J.~F.}\ \bibnamefont {Scott}},\
  }\href@noop {} {\bibfield  {journal} {\bibinfo  {journal} {Nature}\ }\textbf
  {\bibinfo {volume} {442}},\ \bibinfo {pages} {759} (\bibinfo {year}
  {2006})}\BibitemShut {NoStop}%
\bibitem [{\citenamefont {Pisarev}(1994)}]{Pisarev:1994}%
  \BibitemOpen
  \bibfield  {author} {\bibinfo {author} {\bibfnamefont {R.}~\bibnamefont
  {Pisarev}},\ }\href@noop {} {\bibfield  {journal} {\bibinfo  {journal}
  {Ferroelectrics}\ }\textbf {\bibinfo {volume} {162}},\ \bibinfo {pages} {191}
  (\bibinfo {year} {1994})}\BibitemShut {NoStop}%
\bibitem [{\citenamefont {Fiebig}(2005)}]{Fiebig:2005}%
  \BibitemOpen
  \bibfield  {author} {\bibinfo {author} {\bibfnamefont {M.}~\bibnamefont
  {Fiebig}},\ }\href@noop {} {\bibfield  {journal} {\bibinfo  {journal}
  {Journal of Physics D: Applied Physics}\ }\textbf {\bibinfo {volume} {38}},\
  \bibinfo {pages} {R123} (\bibinfo {year} {2005})}\BibitemShut {NoStop}%
\bibitem [{\citenamefont {Ashida}\ \emph {et~al.}(2014)\citenamefont {Ashida},
  \citenamefont {Oida}, \citenamefont {Shimomura}, \citenamefont {Nozaki},
  \citenamefont {Shibata},\ and\ \citenamefont {Sahashi}}]{Achida:2014}%
  \BibitemOpen
  \bibfield  {author} {\bibinfo {author} {\bibfnamefont {T.}~\bibnamefont
  {Ashida}}, \bibinfo {author} {\bibfnamefont {M.}~\bibnamefont {Oida}},
  \bibinfo {author} {\bibfnamefont {N.}~\bibnamefont {Shimomura}}, \bibinfo
  {author} {\bibfnamefont {T.}~\bibnamefont {Nozaki}}, \bibinfo {author}
  {\bibfnamefont {T.}~\bibnamefont {Shibata}}, \ and\ \bibinfo {author}
  {\bibfnamefont {M.}~\bibnamefont {Sahashi}},\ }\href@noop {} {\bibfield
  {journal} {\bibinfo  {journal} {Applied Physics Letters}\ }\textbf {\bibinfo
  {volume} {104}},\ \bibinfo {pages} {152409} (\bibinfo {year}
  {2014})}\BibitemShut {NoStop}%
\bibitem [{\citenamefont {Wilczek}(1987)}]{Wilczek:1987}%
  \BibitemOpen
  \bibfield  {author} {\bibinfo {author} {\bibfnamefont {F.}~\bibnamefont
  {Wilczek}},\ }\href@noop {} {\bibfield  {journal} {\bibinfo  {journal}
  {Physical Review Letters}\ }\textbf {\bibinfo {volume} {58}},\ \bibinfo
  {pages} {1799} (\bibinfo {year} {1987})}\BibitemShut {NoStop}%
\bibitem [{\citenamefont {Li}\ \emph {et~al.}(2010)\citenamefont {Li},
  \citenamefont {Wang}, \citenamefont {Qi},\ and\ \citenamefont
  {Zhang}}]{Li:2010}%
  \BibitemOpen
  \bibfield  {author} {\bibinfo {author} {\bibfnamefont {R.}~\bibnamefont
  {Li}}, \bibinfo {author} {\bibfnamefont {J.}~\bibnamefont {Wang}}, \bibinfo
  {author} {\bibfnamefont {X.-L.}\ \bibnamefont {Qi}}, \ and\ \bibinfo {author}
  {\bibfnamefont {S.-C.}\ \bibnamefont {Zhang}},\ }\href@noop {} {\bibfield
  {journal} {\bibinfo  {journal} {Nature Physics}\ }\textbf {\bibinfo {volume}
  {6}},\ \bibinfo {pages} {284} (\bibinfo {year} {2010})}\BibitemShut {NoStop}%
\bibitem [{\citenamefont {Hasan}\ and\ \citenamefont
  {Kane}(2010)}]{Hasan:2010}%
  \BibitemOpen
  \bibfield  {author} {\bibinfo {author} {\bibfnamefont {M.~Z.}\ \bibnamefont
  {Hasan}}\ and\ \bibinfo {author} {\bibfnamefont {C.~L.}\ \bibnamefont
  {Kane}},\ }\href@noop {} {\bibfield  {journal} {\bibinfo  {journal} {Reviews
  of Modern Physics}\ }\textbf {\bibinfo {volume} {82}},\ \bibinfo {pages}
  {3045} (\bibinfo {year} {2010})}\BibitemShut {NoStop}%
\bibitem [{\citenamefont {Preskill}\ \emph {et~al.}(1983)\citenamefont
  {Preskill}, \citenamefont {Wise},\ and\ \citenamefont
  {Wilczek}}]{Preskill:1983}%
  \BibitemOpen
  \bibfield  {author} {\bibinfo {author} {\bibfnamefont {J.}~\bibnamefont
  {Preskill}}, \bibinfo {author} {\bibfnamefont {M.~B.}\ \bibnamefont {Wise}},
  \ and\ \bibinfo {author} {\bibfnamefont {F.}~\bibnamefont {Wilczek}},\
  }\href@noop {} {\bibfield  {journal} {\bibinfo  {journal} {Physics Letters
  B}\ }\textbf {\bibinfo {volume} {120}},\ \bibinfo {pages} {127} (\bibinfo
  {year} {1983})}\BibitemShut {NoStop}%
\bibitem [{\citenamefont {Popov}\ \emph {et~al.}(1999)\citenamefont {Popov},
  \citenamefont {Kadomtseva}, \citenamefont {Belov}, \citenamefont
  {Vorob�ev},\ and\ \citenamefont {Zvezdin}}]{Popov:1999}%
  \BibitemOpen
  \bibfield  {author} {\bibinfo {author} {\bibfnamefont {Y.~F.}\ \bibnamefont
  {Popov}}, \bibinfo {author} {\bibfnamefont {A.}~\bibnamefont {Kadomtseva}},
  \bibinfo {author} {\bibfnamefont {D.}~\bibnamefont {Belov}}, \bibinfo
  {author} {\bibfnamefont {G.}~\bibnamefont {Vorob�ev}}, \ and\ \bibinfo
  {author} {\bibfnamefont {A.}~\bibnamefont {Zvezdin}},\ }\href@noop {}
  {\bibfield  {journal} {\bibinfo  {journal} {Journal of Experimental and
  Theoretical Physics Letters}\ }\textbf {\bibinfo {volume} {69}},\ \bibinfo
  {pages} {330} (\bibinfo {year} {1999})}\BibitemShut {NoStop}%
\bibitem [{\citenamefont {Fedotov}\ \emph {et~al.}(2013)\citenamefont
  {Fedotov}, \citenamefont {Rogacheva}, \citenamefont {Savinov}, \citenamefont
  {Tsai},\ and\ \citenamefont {Zheludev}}]{Zheludev:2013}%
  \BibitemOpen
  \bibfield  {author} {\bibinfo {author} {\bibfnamefont {V.~A.}\ \bibnamefont
  {Fedotov}}, \bibinfo {author} {\bibfnamefont {A.~V.}\ \bibnamefont
  {Rogacheva}}, \bibinfo {author} {\bibfnamefont {V.}~\bibnamefont {Savinov}},
  \bibinfo {author} {\bibfnamefont {D.~P.}\ \bibnamefont {Tsai}}, \ and\
  \bibinfo {author} {\bibfnamefont {N.~I.}\ \bibnamefont {Zheludev}},\
  }\href@noop {} {\bibfield  {journal} {\bibinfo  {journal} {Scientific
  Reports}\ }\textbf {\bibinfo {volume} {3}},\ \bibinfo {pages} {2967}
  (\bibinfo {year} {2013})}\BibitemShut {NoStop}%
\bibitem [{\citenamefont {Raybould}\ \emph {et~al.}(2016)\citenamefont
  {Raybould}, \citenamefont {Fedotov}, \citenamefont {Papasimakis},
  \citenamefont {Kuprov}, \citenamefont {Youngs}, \citenamefont {Chen},
  \citenamefont {Tsai},\ and\ \citenamefont {Zheludev}}]{Zheludev:2016}%
  \BibitemOpen
  \bibfield  {author} {\bibinfo {author} {\bibfnamefont {T.}~\bibnamefont
  {Raybould}}, \bibinfo {author} {\bibfnamefont {V.}~\bibnamefont {Fedotov}},
  \bibinfo {author} {\bibfnamefont {N.}~\bibnamefont {Papasimakis}}, \bibinfo
  {author} {\bibfnamefont {I.}~\bibnamefont {Kuprov}}, \bibinfo {author}
  {\bibfnamefont {I.}~\bibnamefont {Youngs}}, \bibinfo {author} {\bibfnamefont
  {W.}~\bibnamefont {Chen}}, \bibinfo {author} {\bibfnamefont {D.}~\bibnamefont
  {Tsai}}, \ and\ \bibinfo {author} {\bibfnamefont {N.}~\bibnamefont
  {Zheludev}},\ }\href@noop {} {\bibfield  {journal} {\bibinfo  {journal}
  {Physical Review B}\ }\textbf {\bibinfo {volume} {94}},\ \bibinfo {pages}
  {035119} (\bibinfo {year} {2016})}\BibitemShut {NoStop}%
\bibitem [{\citenamefont {Hehl}\ \emph {et~al.}(2008)\citenamefont {Hehl},
  \citenamefont {Obukhov}, \citenamefont {Rivera},\ and\ \citenamefont
  {Schmid}}]{Obukhov:2008}%
  \BibitemOpen
  \bibfield  {author} {\bibinfo {author} {\bibfnamefont {F.~W.}\ \bibnamefont
  {Hehl}}, \bibinfo {author} {\bibfnamefont {Y.~N.}\ \bibnamefont {Obukhov}},
  \bibinfo {author} {\bibfnamefont {J.-P.}\ \bibnamefont {Rivera}}, \ and\
  \bibinfo {author} {\bibfnamefont {H.}~\bibnamefont {Schmid}},\ }\href@noop {}
  {\bibfield  {journal} {\bibinfo  {journal} {Phys. Rev. A}\ }\textbf {\bibinfo
  {volume} {77}},\ \bibinfo {pages} {022106} (\bibinfo {year}
  {2008})}\BibitemShut {NoStop}%
\bibitem [{\citenamefont {Wiegelmann}\ \emph {et~al.}(1994)\citenamefont
  {Wiegelmann}, \citenamefont {Jansen}, \citenamefont {Wyder}, \citenamefont
  {Rivera},\ and\ \citenamefont {Schmid}}]{Wiegelmann:1994}%
  \BibitemOpen
  \bibfield  {author} {\bibinfo {author} {\bibfnamefont {H.}~\bibnamefont
  {Wiegelmann}}, \bibinfo {author} {\bibfnamefont {A.}~\bibnamefont {Jansen}},
  \bibinfo {author} {\bibfnamefont {P.}~\bibnamefont {Wyder}}, \bibinfo
  {author} {\bibfnamefont {J.-P.}\ \bibnamefont {Rivera}}, \ and\ \bibinfo
  {author} {\bibfnamefont {H.}~\bibnamefont {Schmid}},\ }\href@noop {}
  {\bibfield  {journal} {\bibinfo  {journal} {Ferroelectrics}\ }\textbf
  {\bibinfo {volume} {162}},\ \bibinfo {pages} {141} (\bibinfo {year}
  {1994})}\BibitemShut {NoStop}%
\bibitem [{\citenamefont {Krichevtsov}\ \emph {et~al.}(1993)\citenamefont
  {Krichevtsov}, \citenamefont {Pavlov}, \citenamefont {Pisarev},\ and\
  \citenamefont {Gridnev}}]{Krichevtsov:1993}%
  \BibitemOpen
  \bibfield  {author} {\bibinfo {author} {\bibfnamefont {B.}~\bibnamefont
  {Krichevtsov}}, \bibinfo {author} {\bibfnamefont {V.}~\bibnamefont {Pavlov}},
  \bibinfo {author} {\bibfnamefont {R.}~\bibnamefont {Pisarev}}, \ and\
  \bibinfo {author} {\bibfnamefont {V.}~\bibnamefont {Gridnev}},\ }\href@noop
  {} {\bibfield  {journal} {\bibinfo  {journal} {Journal of Physics: Condensed
  Matter}\ }\textbf {\bibinfo {volume} {5}},\ \bibinfo {pages} {8233} (\bibinfo
  {year} {1993})}\BibitemShut {NoStop}%
\bibitem [{\citenamefont {Foner}\ and\ \citenamefont
  {Hanabusa}(1963)}]{Foner:1963}%
  \BibitemOpen
  \bibfield  {author} {\bibinfo {author} {\bibfnamefont {S.}~\bibnamefont
  {Foner}}\ and\ \bibinfo {author} {\bibfnamefont {M.}~\bibnamefont
  {Hanabusa}},\ }\href@noop {} {\bibfield  {journal} {\bibinfo  {journal}
  {Journal of Applied Physics}\ }\textbf {\bibinfo {volume} {34}},\ \bibinfo
  {pages} {1246} (\bibinfo {year} {1963})}\BibitemShut {NoStop}%
\bibitem [{\citenamefont {Fiebig}\ \emph {et~al.}(1996)\citenamefont {Fiebig},
  \citenamefont {Fr{\"o}hlich},\ and\ \citenamefont {Thiele}}]{Fiebig:1996}%
  \BibitemOpen
  \bibfield  {author} {\bibinfo {author} {\bibfnamefont {M.}~\bibnamefont
  {Fiebig}}, \bibinfo {author} {\bibfnamefont {D.}~\bibnamefont
  {Fr{\"o}hlich}}, \ and\ \bibinfo {author} {\bibfnamefont {H.-J.}\
  \bibnamefont {Thiele}},\ }\href@noop {} {\bibfield  {journal} {\bibinfo
  {journal} {Physical Review B}\ }\textbf {\bibinfo {volume} {54}},\ \bibinfo
  {pages} {R12681} (\bibinfo {year} {1996})}\BibitemShut {NoStop}%
\bibitem [{\citenamefont {Popov}\ \emph {et~al.}(1997)\citenamefont {Popov},
  \citenamefont {Kadomtseva}, \citenamefont {Vorob'Ev}, \citenamefont
  {Timofeeva}, \citenamefont {Tehranchi},\ and\ \citenamefont
  {Zvezdin}}]{popov1997linear}%
  \BibitemOpen
  \bibfield  {author} {\bibinfo {author} {\bibfnamefont {Y.~F.}\ \bibnamefont
  {Popov}}, \bibinfo {author} {\bibfnamefont {A.}~\bibnamefont {Kadomtseva}},
  \bibinfo {author} {\bibfnamefont {G.}~\bibnamefont {Vorob'Ev}}, \bibinfo
  {author} {\bibfnamefont {V.}~\bibnamefont {Timofeeva}}, \bibinfo {author}
  {\bibfnamefont {M.-M.}\ \bibnamefont {Tehranchi}}, \ and\ \bibinfo {author}
  {\bibfnamefont {A.}~\bibnamefont {Zvezdin}},\ }\href@noop {} {\bibfield
  {journal} {\bibinfo  {journal} {Ferroelectrics}\ }\textbf {\bibinfo {volume}
  {204}},\ \bibinfo {pages} {269} (\bibinfo {year} {1997})}\BibitemShut
  {NoStop}%
\bibitem [{\citenamefont {Schmid}(1973)}]{Schmid:1973}%
  \BibitemOpen
  \bibfield  {author} {\bibinfo {author} {\bibfnamefont {H.}~\bibnamefont
  {Schmid}},\ }\href@noop {} {\bibfield  {journal} {\bibinfo  {journal}
  {International Journal of Magnetism}\ }\textbf {\bibinfo {volume} {4}},\
  \bibinfo {pages} {337} (\bibinfo {year} {1973})}\BibitemShut {NoStop}%
\bibitem [{\citenamefont {Gusev}\ \emph {et~al.}(2014)\citenamefont {Gusev},
  \citenamefont {Belotelov},\ and\ \citenamefont {Zvezdin}}]{gusev2014surface}%
  \BibitemOpen
  \bibfield  {author} {\bibinfo {author} {\bibfnamefont {N.}~\bibnamefont
  {Gusev}}, \bibinfo {author} {\bibfnamefont {V.}~\bibnamefont {Belotelov}}, \
  and\ \bibinfo {author} {\bibfnamefont {A.}~\bibnamefont {Zvezdin}},\
  }\href@noop {} {\bibfield  {journal} {\bibinfo  {journal} {Optics letters}\
  }\textbf {\bibinfo {volume} {39}},\ \bibinfo {pages} {4108} (\bibinfo {year}
  {2014})}\BibitemShut {NoStop}%
\bibitem [{\citenamefont {Tse}\ and\ \citenamefont
  {MacDonald}(2010)}]{Wang-Kong:2010}%
  \BibitemOpen
  \bibfield  {author} {\bibinfo {author} {\bibfnamefont {W.-K.}\ \bibnamefont
  {Tse}}\ and\ \bibinfo {author} {\bibfnamefont {A.}~\bibnamefont
  {MacDonald}},\ }\href@noop {} {\bibfield  {journal} {\bibinfo  {journal}
  {Physical Review Letters}\ }\textbf {\bibinfo {volume} {105}},\ \bibinfo
  {pages} {057401} (\bibinfo {year} {2010})}\BibitemShut {NoStop}%
\bibitem [{\citenamefont {Chang}\ and\ \citenamefont
  {Yang}(2009)}]{Chang:2009}%
  \BibitemOpen
  \bibfield  {author} {\bibinfo {author} {\bibfnamefont {M.-C.}\ \bibnamefont
  {Chang}}\ and\ \bibinfo {author} {\bibfnamefont {M.-F.}\ \bibnamefont
  {Yang}},\ }\href@noop {} {\bibfield  {journal} {\bibinfo  {journal} {Physical
  Review B}\ }\textbf {\bibinfo {volume} {80}},\ \bibinfo {pages} {113304}
  (\bibinfo {year} {2009})}\BibitemShut {NoStop}%
\bibitem [{\citenamefont {Liu}\ \emph {et~al.}(2014)\citenamefont {Liu},
  \citenamefont {Xu},\ and\ \citenamefont {Yang}}]{Liu:2014}%
  \BibitemOpen
  \bibfield  {author} {\bibinfo {author} {\bibfnamefont {F.}~\bibnamefont
  {Liu}}, \bibinfo {author} {\bibfnamefont {J.}~\bibnamefont {Xu}}, \ and\
  \bibinfo {author} {\bibfnamefont {Y.}~\bibnamefont {Yang}},\ }\href@noop {}
  {\bibfield  {journal} {\bibinfo  {journal} {JOSA B}\ }\textbf {\bibinfo
  {volume} {31}},\ \bibinfo {pages} {735} (\bibinfo {year} {2014})}\BibitemShut
  {NoStop}%
\bibitem [{\citenamefont {Kalish}\ \emph {et~al.}(2007)\citenamefont {Kalish},
  \citenamefont {Belotelov},\ and\ \citenamefont {Zvezdin}}]{Kalish:2007}%
  \BibitemOpen
  \bibfield  {author} {\bibinfo {author} {\bibfnamefont {A.~N.}\ \bibnamefont
  {Kalish}}, \bibinfo {author} {\bibfnamefont {V.~I.}\ \bibnamefont
  {Belotelov}}, \ and\ \bibinfo {author} {\bibfnamefont {A.~K.}\ \bibnamefont
  {Zvezdin}},\ }\href {\doibase 10.1117/12.752467} {\bibfield  {journal}
  {\bibinfo  {journal} {Proc. SPIE}\ }\textbf {\bibinfo {volume} {6728}},\
  \bibinfo {pages} {67283D} (\bibinfo {year} {2007})}\BibitemShut {NoStop}%
\bibitem [{\citenamefont {Liscidini}\ and\ \citenamefont
  {Sipe}(2010)}]{Liscidini:2010}%
  \BibitemOpen
  \bibfield  {author} {\bibinfo {author} {\bibfnamefont {M.}~\bibnamefont
  {Liscidini}}\ and\ \bibinfo {author} {\bibfnamefont {J.}~\bibnamefont
  {Sipe}},\ }\href@noop {} {\bibfield  {journal} {\bibinfo  {journal} {Physical
  Review B}\ }\textbf {\bibinfo {volume} {81}},\ \bibinfo {pages} {115335}
  (\bibinfo {year} {2010})}\BibitemShut {NoStop}%
\bibitem [{\citenamefont {Belotelov}\ \emph {et~al.}(2011)\citenamefont
  {Belotelov}, \citenamefont {Akimov}, \citenamefont {Pohl}, \citenamefont
  {Kotov}, \citenamefont {Kasture}, \citenamefont {Vengurlekar}, \citenamefont
  {Gopal}, \citenamefont {Yakovlev}, \citenamefont {Zvezdin},\ and\
  \citenamefont {Bayer}}]{TMOKE_in_PlC}%
  \BibitemOpen
  \bibfield  {author} {\bibinfo {author} {\bibfnamefont {V.}~\bibnamefont
  {Belotelov}}, \bibinfo {author} {\bibfnamefont {I.}~\bibnamefont {Akimov}},
  \bibinfo {author} {\bibfnamefont {M.}~\bibnamefont {Pohl}}, \bibinfo {author}
  {\bibfnamefont {V.}~\bibnamefont {Kotov}}, \bibinfo {author} {\bibfnamefont
  {S.}~\bibnamefont {Kasture}}, \bibinfo {author} {\bibfnamefont
  {A.}~\bibnamefont {Vengurlekar}}, \bibinfo {author} {\bibfnamefont {A.~V.}\
  \bibnamefont {Gopal}}, \bibinfo {author} {\bibfnamefont {D.}~\bibnamefont
  {Yakovlev}}, \bibinfo {author} {\bibfnamefont {A.}~\bibnamefont {Zvezdin}}, \
  and\ \bibinfo {author} {\bibfnamefont {M.}~\bibnamefont {Bayer}},\
  }\href@noop {} {\bibfield  {journal} {\bibinfo  {journal} {Nature
  Nanotechnology}\ }\textbf {\bibinfo {volume} {6}},\ \bibinfo {pages} {370}
  (\bibinfo {year} {2011})}\BibitemShut {NoStop}%
\end{thebibliography}%
\end{document}